\documentclass[runningheads,envcountsame]{llncs}

\makeatletter \RequirePackage[bookmarks,unicode,colorlinks=true]{hyperref}%
\def\@citecolor{blue}%
\def\@urlcolor{blue}%
\def\@linkcolor{blue}%

\def\orcidID#1{\href{http://orcid.org/#1}{\smash{\protect\raisebox{-1.25pt}{\protect\includegraphics{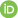}}}}}
\makeatother

\usepackage{mathtools,amssymb,amsfonts}
\usepackage{tikz}
\usetikzlibrary{shapes,calc,arrows,automata,decorations.pathmorphing,decorations.pathreplacing}
\usepackage{marvosym}
\usepackage[capitalize,nameinlink]{cleveref}
\usepackage{subcaption}
\usepackage{tabto}
\usepackage{xspace}
\usepackage{booktabs}
\usepackage{tabto}

\newenvironment{myproof}{
  \noindent{\it Proof.}
}{\qed
  \medskip
}
\usepackage[location=appendix,prependtoappendix=true,hideproofs=false]{moveproofs}
\usepackage{thmtools, thm-restate}
\usepackage{xcolor}
\newcommand{\tool}[1]{\textsf{#1}}
\newcommand{\KoAT}[0]{\tool{KoAT}}
\newcommand{\set}[1]{\left\lbrace #1 \right\rbrace}
\newcommand{\var}{\normalfont\texttt}
\newcommand{\wildcard}{\underline{\hspace{0.15cm}}}
\renewcommand{\epsilon}{\varepsilon}

\newcommand{\NN}{\mathbb{N}}
\newcommand{\ZZ}{\mathbb{Z}}
\newcommand{\QQ}{\mathbb{Q}}
\newcommand{\RR}{\mathbb{R}}

\newcommand{\NNC}{\overline{\mathbb{N}}}

\renewcommand{\emptyset}{\varnothing}

\newcommand{\abs}[1]{\lvert #1 \rvert}
\renewcommand{\approx}[1]{\lceil\!\!\lceil #1 \rceil\!\!\rceil}
\newcommand{\landau}{\mathcal{O}}

\newcommand{\tree}{\mathbb{T}}

\newcommand{\AtomSet}{\mathcal{A}}
\newcommand{\ConstraintSet}{\mathcal{C}}

\newcommand{\true}{\var{true}}

\newcommand{\initial}{\sigma_0}
\newcommand{\valuation}{\sigma}
\newcommand{\Valuation}{\Sigma}

\newcommand{\VSet}{\mathcal{V}}
\newcommand{\indv}{d}
\newcommand{\guard}{\varphi}
\newcommand{\update}{\eta}
\newcommand{\recupdate}{\zeta}

\newcommand{\TSet}{\mathcal{T}}
\newcommand{\LSet}{\mathcal{L}}
\newcommand{\RetSet}{\Omega}
\newcommand{\FSet}{\mathcal{F}}

\newcommand{\RVSet}{\mathcal{RV}}
\newcommand{\location}{\ell}
\newcommand{\IntProgram}{(\LSet,\location_0,\RetSet,\FSet,\TSet)}
\DeclareMathOperator{\rc}{rc}

\newcommand{\entryT}{\mathcal{ET}}
\newcommand{\entryF}{\mathcal{EF}}

\newcommand{\step}{\prec}
\DeclareMathOperator{\fun}{fun}
\DeclareMathOperator{\trans}{trans}
\newcommand{\its}{$\rho$-ITS}
\newcommand{\rrf}{$\rho$-RF}
\DeclareMathOperator{\nfc}{nfc}

\newcommand{\pret}{r}

\newcommand{\actstate}{{\tilde{\valuation}}}

\DeclareMathOperator{\actV}{actV}
\DeclareMathOperator{\actF}{actF}
\DeclareMathOperator{\actA}{act}
\DeclareMathOperator{\pre}{pre}
\newcommand{\rv}[1]{\langle #1 \rangle}

\usepackage{stmaryrd}
\newcommand{\eval}[2]{\llbracket #1 \rrbracket_{#2}}

\newcommand{\BoundSet}{\mathcal{B}}
\newcommand{\Size}{{\mathcal{SB}}}
\newcommand{\SizeLoc}{{\mathcal{SB}_{\normalfont\text{loc}}}}

\newcommand{\UTime}{{\mathcal{RB}}}
\newcommand{\loc}{\UTime_{\normalfont\text{loc}}}
\newcommand{\locParameter}[2]{{\UTime_{\,\normalfont\text{loc}}^{\,#1,#2}}}
\newcommand{\glo}{\mathcal{RB}}
\newcommand{\glopr}{\mathcal{RB}'}

\crefname{definition}{Def.}{Def.}
\crefname{example}{Ex.}{Ex.}
\crefname{counterexample}{Counterex.}{Counterex.}
\crefname{appendix}{App.}{App.}
\crefname{ex}{Ex.}{Ex.}
\crefname{theorem}{Thm.}{Thm.}
\crefname{lemma}{Lemma}{Lemmas}
\crefname{remark}{Rem.}{Rem.}
\crefname{section}{Sect.}{Sect.}
\crefname{subsection}{Sect.}{Sect.}
\crefname{subsubsection}{Sect.}{Sect.}
\crefname{line}{Line}{Lines}
\crefname{corollary}{Cor.}{Cor.}
\crefname{figure}{Fig.}{Fig.}
\crefname{enumi}{}{}
\crefname{algorithm}{Alg.}{Alg.}

\newcommand{\cl}[1]{\texttt{cl}_{#1}}
\newcommand{\PPEE}{\mathbb{PE}}
\newcommand{\lex}{>_{\mathrm{lex}}}

\newcommand{\retVar}{\mathit{retvar}}

\usepackage[hyperref=true, backend=bibtex, firstinits=true, maxbibnames=99, sortcites, style=numeric-comp]{biblatex}
\hypersetup{
  pdftitle={Modular Automatic Complexity Analysis of Recursive Integer Programs}, colorlinks=true, linkcolor=blue, citecolor=olive, filecolor=magenta, urlcolor=cyan
}
\usepackage[shortlabels]{enumitem}
\setlist[enumerate,1]{label=(\alph*), wide=0pt, leftmargin=*}
\setlist[itemize,1]{label=\textbullet}

\newcommand{\paper}[1]{}
\newcommand{\report}[1]{#1}

\report{
\usepackage{xurl}
\hypersetup{breaklinks=true}
}

\author{Nils Lommen$^{(\href{mailto:lommen@cs.rwth-aachen.de}{\mbox{\Letter}})}$\orcidID{0000-0003-3187-9217} \and Jürgen Giesl\orcidID{0000-0003-0283-8520}}
\authorrunning{N.\ Lommen \and J.\ Giesl}
\institute{RWTH Aachen University, Aachen, Germany\\
  \email{\{lommen,giesl\}@cs.rwth-aachen.de}}
\title{Modular Automatic Complexity Analysis of Recursive Integer Programs}
\titlerunning{Modular Automatic Complexity Analysis of Recursive Integer Programs}

\allowdisplaybreaks

\begin{document}
\maketitle
\begin{abstract}
  In earlier work, we developed a modular approach for automatic complexity analysis of integer programs.
  However, these integer programs do not allow non-tail \emph{recursive} calls or subprocedures.
  In this paper, we consider integer programs with function calls and present a natural extension of our modular complexity analysis approach to the recursive setting based on a new form of ranking functions.
  Hence, our approach combines already existing powerful techniques on the ``imperative'' parts of the program and our novel ranking functions on the recursive parts.
  The strength of this combination is demonstrated by our implementation in the complexity analysis tool \KoAT{}.
\end{abstract}

\section{Introduction}
There exist numerous approaches to analyze complexity of programs automatically, e.g., \cite{ben-amram2017MultiphaseLinearRankingFunctions,albert2019ResourceAnalysisDriven,sinn2017ComplexityResourceBound,brockschmidt2016AnalyzingRuntimeSize,Flores-MontoyaH14,giesl2022ImprovingAutomaticComplexity,hoffmann2017AutomaticResourceBound,lopez18IntervalBasedResource,carbonneaux2015CompositionalCertifiedResource,albert2012CostAnalysisObjectoriented,lommen2024ControlFlowRefinementProbabilistic,pham2024RobustResourceBounds,hoffmann2012MultivariateAmortizedResource,albert2013InferenceResourceUsage,handley2019LiquidateYourAssets}, but most of them are essentially limited to non-recursive programs.
There are also several techniques for complexity analysis of term rewrite systems (TRSs) \cite{baader_nipkow_1998} which can handle arbitrary recursion.
However, TRSs have the drawback that they do not support built-in data types like integers.
Thus, the goal of this paper is to automatically analyze the complexity of programs with built-in integers \emph{and} arbitrary (possibly non-tail) recursion.

In previous work, we developed a \emph{modular} technique for complexity analysis of programs with built-in integers which we implemented in the complexity analysis tool \KoAT{} \cite{brockschmidt2016AnalyzingRuntimeSize,giesl2022ImprovingAutomaticComplexity,lommen2022AutomaticComplexityAnalysis,lommen2023TargetingCompletenessUsing,lommen2024TargetingCompletenessComplexity}.
It automatically infers runtime bounds for integer transition systems (ITSs) possibly consisting of multiple loops by handling some subprograms as \emph{twn}-loops (where there exist ``complete'' techniques for analyzing termination and complexity \cite{lommen2022AutomaticComplexityAnalysis,lommen2023TargetingCompletenessUsing,lommen2024TargetingCompletenessComplexity,frohn2020TerminationPolynomialLoops,hark2020PolynomialLoopsTermination}) and by using multiphase-linear ranking functions \cite{giesl2022ImprovingAutomaticComplexity,ben-amram2017MultiphaseLinearRankingFunctions,ben-amram2019MultiphaseLinearRankingFunctions}
for other subprograms.
By inferring bounds for one subprogram after the other, in the end we obtain a bound on the runtime of the whole program.
In this paper, we extend our approach to ITSs which allow \emph{function calls}, including non-tail recursion.
In the first attempt for such an extension from \cite[Sect.\ 5]{brockschmidt2016AnalyzingRuntimeSize}, the results of function calls were simply disregarded.
In contrast, our novel approach can take the results of function calls into account which leads to a much higher precision.
\begin{figure}[t]
  \begin{minipage}[t]{0.5\textwidth}
    \begin{align*}
       & \textbf{main($x,y$):} \\
       & \textbf{while $x > 0$ do} \\
       & \quad y \gets y + \textbf{fac($x$)};\; x \gets x - 1; \\
       & x \gets 1; \\
       & \textbf{while $x < y$ do} \\
       & \quad x \gets 3 \cdot x; \; y \gets 2 \cdot y;
    \end{align*}
  \end{minipage}
  \begin{minipage}[t]{0.4\textwidth}
    \begin{align}
       & \textbf{fac($a$):} \nonumber \\
       & \textbf{if $a = 0$ then } \nonumber \\
       & \quad \textbf{return } 1; \nonumber \\
       & \textbf{else if $a > 0$ then} \nonumber \\
       & \quad \textbf{return } a\cdot\textbf{fac($a - 1$)}; \nonumber
    \end{align}
  \end{minipage}
  \caption{Recursive Integer Program with two Procedures}\label{fig:IntProg}
\end{figure}

\begin{example}
  \label{exa:program}
  The first \textbf{while}-loop of the procedure \textbf{main} in \Cref{fig:IntProg}
  computes $x! + \linebreak
    \dots + 1!$ by calling the subprocedure \textbf{fac}. We introduce a novel class of ranking functions for recursive programs to show that this loop has quadratic runtime (when every assignment has the ``cost'' 1).
  Note that $y$'s value is bounded by $y + x\cdot x^x$, where $x$ and $y$ refer to the initial values of the program variables.
  The reason is that $x! + \dots + 1!$ can be over-approximated by $x\cdot x^x$ since $1!,\dots,x!$ are all bounded by $x^x$.
  This observation is crucial for the runtime of the second loop since it is executed at most $\log_2(\text{size}(y)) + 2 = \log_2(y + x \cdot x^x) + 2$ times, where ``$\text{size}(y)$'' denotes such an over-approximation of the (absolute) value of $y$ before the second loop.
  Hence, as the runtime of the first loop is quadratric and the runtime of the second loop is less than quadratic, the overall program has quadratic runtime.
  Here, \cite[Sect.\ 5]{brockschmidt2016AnalyzingRuntimeSize} fails to infer a finite runtime bound, as it disregards the return value of \textbf{fac}.
  The runtime bound $\log_2(y) + 2$ for the second loop can be obtained by our technique based on \emph{twn}-loops, but not by linear ranking functions.
  Thus, our novel approach for recursive integer programs allows us, e.g., to apply ranking functions on some (possibly recursive) parts of the program and techniques for \emph{twn}-loops on other parts, i.e., it allows for modular proofs that use different techniques for automated complexity analysis on different subprograms in order to benefit from their individual strengths.
\end{example}

\vspace*{-.1cm}

In this work, we extend our notions of runtime and size bounds \cite{brockschmidt2016AnalyzingRuntimeSize,giesl2022ImprovingAutomaticComplexity,lommen2022AutomaticComplexityAnalysis,lommen2023TargetingCompletenessUsing,lommen2024TargetingCompletenessComplexity} to the new setting of ITSs with function calls.
On the one hand, as illustrated by \Cref{exa:program}, we need size bounds to compute runtime bounds.
On the other hand, we also need runtime bounds to infer size bounds, because to this end we have to over-approximate how often loops with variable updates are iterated.
Thus, our approach alternates between the computation of runtime and size bounds.
All our contributions are implemented in our complexity analysis tool \KoAT{}.

\paragraph{Structure:}
In \Cref{sect:recITS} we introduce our new notion of ITSs with function calls and define runtime and size bounds for these programs.
In \Cref{sect:RB}, we show how to compute modular runtime bounds for our new class of programs.
Analogously, we present a technique to infer size bounds in a modular way in \Cref{sect:SB}.
Finally, in \Cref{sect:conclusion} we discuss related work and our implementation in the tool \KoAT{}, and provide an experimental evaluation demonstrating the strengths of our approach.
All proofs can be found in\report{ App.\ A.}\paper{ \cite[App.\ A]{report}.}

\section{Recursive Integer Transition Systems}
\label{sect:recITS}

In \Cref{Syntax and Semantics} we extend ITSs by function calls and recursion.
Afterwards, in \Cref{Runtime and Size Bounds} we define runtime and size bounds which extend the corresponding notions for ITSs without function calls \cite{brockschmidt2016AnalyzingRuntimeSize,giesl2022ImprovingAutomaticComplexity,lommen2022AutomaticComplexityAnalysis,lommen2023TargetingCompletenessUsing,lommen2024TargetingCompletenessComplexity}
in a natural way.

\subsection{Syntax and Semantics of Recursive Integer Transition Systems}
\label{Syntax and Semantics}

We fix a finite set of program variables $\VSet$.
As usual, $\ZZ[\VSet]$ is the polynomial ring over the vari\-ables $\VSet$ with integer coefficients.
We use polynomials for the \emph{constraints} in the guards of transitions.

\begin{definition}[Atoms and Constraints]
  \label{def:formulas}
  The set of \emph{atoms} $\AtomSet$ consists of all inequations $p_1 < p_2$ for polynomials $p_1,p_2\in\ZZ[\VSet]$.
  The set $\ConstraintSet$ of \emph{constraints} consists of all formulas built from atoms $\AtomSet$ and $\land$.
\end{definition}
We also use ``$\geq$'', ``$=$'', ``$\neq$'', and negations ``$\neg$'', since they can be simulated by atoms and constraints (e.g., $p_1 \geq p_2$ is equivalent to $p_2 < p_1 + 1$ for integers).
Disjunctions ``$\lor$'' are modeled by several transitions with the same start and target location.

ITSs are a widely studied formalism in automatic program verification (see, e.g., \cite{dingel1995ModelCheckingInfinite,henzinger1995ComputingSimulationsFinite,abdulla1996GeneralDecidabilityTheorems,kupferman2000AutomataTheoreticApproachReasoning,brockschmidt2016AnalyzingRuntimeSize,chatterjee2025SoundCompleteWitnesses}).
An ITS consists of a set $\LSet$ of \emph{locations} and a set $\TSet$ of \emph{transitions}, where a transition connects two locations.
Moreover, every transition has a polynomial \emph{update} function $\update : \VSet\to\ZZ[\VSet]$.
The values of the variables are ``stored'' in a \emph{state} $\valuation: \VSet \rightarrow \ZZ$, where $\Valuation$ denotes the set of all states.
When evaluating a transition, the values of the variables are changed according to its update function, and we move from a \emph{configuration} $(\location,\valuation)\in\LSet\times\Valuation$ to another configuration $(\location',\valuation')$.
Since ITSs do not allow any non-tail recursion, \Cref{def:recITS} extends them to ITSs with \emph{function calls} (called \its{s}).
To this end, we introduce a set $\FSet$ of \emph{function calls} $\location(\recupdate)$ which can now occur in the updates of variables as well.
Here, $\location$ is the start location of the subprogram that is called and the update $\recupdate: \VSet \to \ZZ[\VSet]$ sets the program variables of the subprogram to their initial values.
More precisely, if $\valuation\in\Valuation$ is the state before calling the subprogram via $\location(\recupdate)$, then the subprogram starts in a configuration $(\location,\tilde{\valuation})$, where $\tilde{\valuation}$ results from ``applying'' the update $\recupdate$ to the state $\valuation$.
For example, if $\recupdate_1(a) = x$, then $f_1(\recupdate_1)$ represents the function call that sets $a$ to the current value of $x$ and jumps to the location $f_1$.

In addition, we also introduce a set of \emph{return locations} $\RetSet \subseteq \LSet$, and for every return location $\location' \in \RetSet$, we let $v_{\location'} \in \VSet$ denote its \emph{return variable}.
As soon as a called subprogram reaches a configuration $(\location',\valuation')$ with $\location' \in\RetSet$, the value $\valuation'(v_{\location'})$ is returned as the result of the function call $\location(\recupdate)$.
(We will define the semantics of \its{s} formally in \Cref{def:semantic_recITS,def:semantic_recITS_pop}.)
Thus, transitions may now have updates which map variables to polynomial combinations of variables \emph{and} function calls.
This is denoted by $\ZZ[\VSet\cup\FSet]$.
In this way, the results of function calls can be combined polynomially, and they can be used by transitions when updating variables.

\begin{definition}[\its{}]
  \label{def:recITS}
  The tuple $\IntProgram$ is an \emph{ITS with function calls} (\its{}) where
  \begin{itemize}
    \item $\LSet$ is a finite set of \emph{locations} with an initial location $\location_0 \in \LSet$,
    \item $\FSet$ is a finite set of \emph{function calls} $\location(\recupdate)$ with $\location\in\LSet\setminus\set{\location_0}$ and $\recupdate: \VSet \to \ZZ[\VSet]$,
    \item $\RetSet\subseteq\LSet$ is a set of \emph{return locations}
          and for every $\location \in \RetSet$, $v_{\location} \in \VSet$ denotes its \emph{return variable}, and
    \item $\TSet$ is a finite set of \emph{transitions}: A transition is a 4-tuple $(\location,\guard,\update,\location')$ with start location $\location\in\LSet$, target loca\-tion $\location'\in\LSet\setminus\set{\location_0}$, guard $\guard\in\ConstraintSet$, and update function $\update:\VSet\to\ZZ[\VSet\cup\FSet]$.
  \end{itemize}
\end{definition}
We denote the set of function calls in a polynomial $p$, an update $\update$, or a tran\-sition $t$ by $\fun(p)$, $\fun(\update)$, or $\fun(t)$, respectively.
Similarly, for a function call $\rho \in \FSet$, $\trans(\rho)$ is the set of all transitions of the ITS in which $\rho$ occurs in an update.

Note that our definition of \its{s} allows
non-deterministic branching since several transitions can have the same start location.
Moreover, to model non-deterministic sampling, our approach can easily be extended by additional ``temporary'' variables which are updated arbitrarily in each evaluation step.
Intuitively, these variables are set non-deterministically by an adversary trying to ``sabotage'' the program in order to obtain long runtimes.
However, we omitted such temporary variables from the paper to simplify the presentation.

\begin{example}
  \label{exa:its}
  The \its{} in \Cref{fig:ITS} corresponds to the program from \Cref{fig:IntProg}.
  In \Cref{fig:ITS}, we omitted trivial guards $\guard = \true$ and identity updates of the form $\update(v) = v$.
  The \its{} has the program variables $\VSet = \set{a,x,y}$, five locations $\LSet = \{\location_0,\location_1, \location_2, f_1,f_2\}$, and two function calls $\rho_1 = f_1(\recupdate_1)$ and $\rho_2 = f_1(\recupdate_2)$, where $\recupdate_1(a) = x$ and $\recupdate_2(a) = a-1$, and both $\recupdate_1$ and $\recupdate_2$ keep $x$ and $y$ unchanged.
  To ease readability, we wrote $f_1(x,x,y)$ and $f_1(a-1,x,y)$ instead of $f_1(\recupdate_1)$ and $f_1(\recupdate_2)$ in \Cref{fig:ITS}.
  The subprogram with the locations $f_1$ and $f_2$ computes the factorial $a!$ recursively and returns this result in the return variable $a$ when reaching the return location $f_2$ (indicated by the doubled node).
  This subprogram is called iteratively in the loop $t_1$ with the argument $x$ (i.e., with $\recupdate_1$ where $\recupdate_1(a) = x$).
  The factorials $x!, (x-1)!, \ldots, 1$ are summed up in the variable $y$.
  Afterwards, $x$ is set to $1$ in $t_2$, and the second loop $t_3$ at location $\location_2$ is executed.
  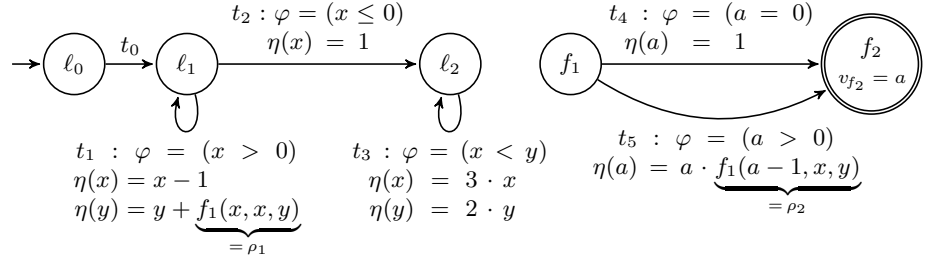
\begin{figure}[t]
    \centering \hspace*{-0.4cm}
    \begin{tikzpicture}[->,>=stealth',shorten >=1pt,auto,node distance=3.5cm,semithick,initial text=$ $]
      \node[state,initial] (q0) {$\location_0$};
      \node[state] (q1) [right of=q0,xshift=-2cm]{$\location_1$};
      \node[state] (q2) [right of=q1, node distance=3.5cm]{$\location_2$};
      \node[state] (q3) [right of=q2, node distance=1.6cm]{$f_1$};
      \node[state,accepting] (q4) [right of=q3, node distance=4cm,text width=.7cm,align=center]{$f_2$\\
      {\scriptsize \hspace*{-.2cm} ${v\!_{f_2} = a}$}};
      \draw (q0) edge node [text width=3.5cm,align=center,above] {\footnotesize $t_0$} (q1);
      \draw (q1) edge [loop below] node [text width=4.7cm,align=center,below]
        {\footnotesize $t_1: \guard = (x > 0)$\\
          $
            \begin{array}{rcl}
              \update(x) & = & x - 1                                   \\
              \update(y) & = & y + \underbrace{f_1(x,x,y)}_{=\,\rho_1} \\
            \end{array}
          $
        } (q1);
      \draw (q1) edge node [text width=2.5cm,align=center,above] {$t_2: \guard = (x \leq 0)$ \\
          $\update(x) = 1$} (q2);
      \draw (q2) edge [loop below] node [text width=3.5cm,align=center,below] {\footnotesize $t_3: \guard = (x < y)$\\
          $\hspace{-0.2cm}\update(x) = 3\cdot x$\\
          $\hspace{-0.2cm}\update(y) = 2\cdot y$} (q2);
      \draw (q3) edge node [text width=4cm,align=center,above] {$t_4: \guard = (a = 0)$\\
          $\hspace*{-0.7cm}\update(a) = 1$} (q4);
      \draw (q3) edge [bend right] node [text width=4.5cm,align=center,below,xshift=.2cm]
        {$t_5: \guard = (a > 0)$\\
          $\update(a) = a \cdot \underbrace{f_1(a - 1,x,y)}_{=\,\rho_2}$
        } (q4);
    \end{tikzpicture}
    \caption{An Integer Transition System with Function Calls $\rho_1$ and $\rho_2$}\label{fig:ITS}
  \end{figure}
\end{example}

Let $\eval{e}{\valuation}$ denote the \emph{evaluation} of an expression $e$ in the state $\valuation\in\Valuation$, where\linebreak
$\eval{e}{\valuation}$ results from replacing every variable $v$ in $e$ by its value $\valuation(v)$.
So, for example, evaluating $\eval{3\cdot x}{\valuation}$ and $\eval{x > 0}{\valuation}$ at $\valuation(x) = 2$ results in $6$ and $\true$, respectively.

From now on, we fix a \its{} $\IntProgram$ over the variables $\VSet$.
Formal\-ly, an evaluation step of a \its{} is a transformation of an \emph{evaluation tree} $\tree$ whose nodes are labeled with configurations from $\LSet\times(\Valuation\cup\set{\bot})$ and whose edges are labeled with transitions or function calls.
We distinguish two kinds of evaluation steps: $t$-evaluation steps (for transitions $t \in \TSet$) and $\epsilon$-evaluation steps.

If a leaf of $\tree$ is labeled with a configuration $(\location,\valuation)$ where a transition $t = (\location,\guard,\update,\location')$ can be applied, then a \emph{$t$-evaluation step} extends $\tree$ at the position of this leaf to a new tree $\tree'$, denoted $\tree \step_t \tree'$.
If the update $\update$ does not contain any function calls, then $\tree'$ results from $\tree$ by adding an edge from the node labeled\linebreak
with $(\location,\valuation)$ to a new node labeled with a configuration $(\location',\valuation')$ where $\valuation'(v) = \eval{\update(v)}{\valuation}$ for all program variables $v$.
This new edge is labeled with $t$, i.e., $(\location,\valuation) \to_t (\location',\valuation')$.
This corresponds to ordinary evaluations of ITSs as in \cite{lommen2023TargetingCompletenessUsing,lommen2024TargetingCompletenessComplexity,giesl2022ImprovingAutomaticComplexity}.

However, if $t$'s update contains function calls $\rho_i = \location_i(\recupdate_i)$ for $1 \leq i \leq n$, then $\tree'$ results from $\tree$ by adding $n+1$ children to the former leaf labeled with $(\location,\valuation)$: The child $(\location',\bot)$ is connected by an edge labeled with $t$, i.e., $(\location,\valuation) \to_t (\location',\bot)$.
Here, $\bot$ denotes an undefined state which will be instantiated later if the function calls $\rho_1, \ldots, \rho_n$ reach return locations.
Moreover, the children $(\location_i,\valuation_i)$ are connected by edges labeled with $\rho_i$ for all $1 \leq i \leq n$, i.e., $(\location,\valuation) \to_{\rho_i} (\location_i,\valuation_i)$, where $\valuation_i(v) = \eval{\recupdate_i(v)}{\valuation}$ for all program variables $v$.

When modeling the semantics in an alternative stack-based way, the $\rho_i$-edges would correspond to a \textsf{push}-operation where the function call $\rho_i$ is pushed on top of the call stack.
Our evaluation trees are an explicit representation of such stacks.
They lift the semantics of ITSs to \its{s} in a natural way by moving from\linebreak
evaluation \emph{paths} to \emph{trees}.
An evaluation tree keeps track of every state that was reached during the evaluation.
While our operational semantics via evaluation trees are equivalent to the stack-based semantics of recursion, the advantage over the stack representation is that evaluation trees are particularly suitable for our novel \emph{$\rho$-ranking functions} for transitions with function calls in \Cref{sec:Ranking Functions}.

For an initial state $\initial\in\Valuation$, the evaluation always starts with the initial evaluation tree $\tree_{\initial} = (\set{(\location_0,\initial)},\emptyset)$ which consists of the single node $(\location_0,\initial)$ and does not have any edges.
We now define how to extend such trees using $t$- and $\epsilon$-steps.
Then the set of \emph{evaluation trees}
is the smallest set of trees that can be obtained from such initial trees by repeated application of $t$- and $\epsilon$-steps.

\begin{definition}[Evaluation of \its{s} ($t$-Evaluation Step)]
  \label{def:semantic_recITS}
  Let $\tree$ be an evaluation tree.
  $\tree \step_t \tree'$ is a \emph{$t$-evaluation step} with the tran\-sition $t =( \location,\guard,\update,\location')$ iff $\,\tree$ has a leaf labeled with $(\location,\valuation)$ where $\valuation \in \Valuation$, $\eval{\guard}{\valuation} = \true$, and

  \begin{itemize}
    \item if $\fun(\update) = \emptyset$, then $\tree'$ is the extension of $\tree$ by an edge $(\location,\valuation) \to_t (\location',\valuation')$ to a new node labeled with $(\location',\valuation')$ where $\valuation'(v) = \eval{\update(v)}{\valuation}$ for all $v\in\VSet$.
    \item if $\update$ contains the function calls $\rho_1 = \location_1(\recupdate_1),\dots,\rho_n = \location_n(\recupdate_n)$, then $\tree'$ is the extension of $\tree$ by the edges $(\location,\valuation) \to_t (\location',\bot)$ and $(\location,\valuation) \to_{\rho_i} (\location_i,\valuation_i)$, where $\valuation_i(v) = \eval{\recupdate_i(v)}{\valuation}$ for all $v\in\VSet$ and all $1 \leq i \leq n$.
  \end{itemize}
\end{definition}

To instantiate the undefined state $\bot$, we use \emph{$\epsilon$-evaluation steps}.
An $\epsilon$-evalua\-tion step corresponds to a \textsf{pop}-operation in the stack-based semantics, i.e., it is used to return the value of a function call after it has been fully evaluated.
If the tree contains $(\location,\valuation) \to_t (\location',\bot)$ and $(\location,\valuation) \to_{\rho_i} (\location_i,\valuation_i)$ for $1 \leq i \leq n$, and there are paths from the nodes $(\location_i,\valuation_i)$ to configurations $(\location_i',\valuation_i')$ where $\location_i'\in\RetSet$ is a return location, then the undefined state $\bot$ can be replaced by a state $\valuation'$ such that $\valuation'(v) = \eval{\overline{\update}(v)}{\valuation}$ for all program variables $v$.
If $\update$ is the update of the transition $t$, then $\overline{\update}(v)$ results from $\update(v)$ by replacing every function call $\location_i(\recupdate_i)$ occurring in $\update$ by the corresponding returned value $\valuation_i'(v_{\location_i'})$ for all $1 \leq i \leq n$.
We denote this by $\overline{\update}(v) = \update(v)\, [\location_i(\recupdate_i)/\valuation_i'(v_{\location_i'})]$.
Thus, an $\epsilon$-evaluation step does not add new edges to an evaluation tree; it merely replaces $\bot$ by a state with concrete values obtained from the evaluated function calls.
\begin{definition}[Evaluation of \its{s} ($\epsilon$-Evaluation Step)]
  \label{def:semantic_recITS_pop}
  Let $\tree$ be an evaluation tree.
  Furthermore, let there be a transition $t = (\location,\guard,\update,\location')$ such that $\tree$ contains a node labeled with $(\location,\valuation)$, with edges and children nodes of the form $(\location,\valuation) \to_t (\location',\bot)$ and $(\location,\valuation) \to_{\rho_i} (\location_i,\valuation_i)$ for function calls $\rho_i = \location_i(\recupdate_i)\in\fun(\update)$.

  If $\tree$ contains paths from each child labeled with $(\location_i,\valuation_i)$ to a node labeled with $(\location_i',\valuation_i')\in \RetSet\times\Valuation$, then $\tree \step_\epsilon \tree'$ is an \emph{$\epsilon$-evaluation step} iff $\tree'$ results from $\tree$ by replacing the label $(\location',\bot)$ by $(\location',\valuation')$, where $\valuation'(v) = \eval{\update(v)\,[\location_i(\recupdate_i)/\valuation_i'(v_{\location_i'})]}{\valuation}$ for all variables $v\in\VSet$.
\end{definition}

We write $\step_{\TSet}$ for $\bigcup_{t \in \TSet} \step_{t}$ and $\step$ for $\step_{\TSet \cup \{ \epsilon \}}$.
Moreover, we denote finitely many evaluations steps $\tree\step\cdots\step \tree'$ by $\tree\step^* \tree'$.

\begin{example}
  \label{exa:evaluation}
  \begin{figure}[t]
    \begin{tikzpicture}[->,>=stealth',shorten >=1pt,auto,state without output/.append style={rectangle,minimum size=0pt,draw=none}]
      \node[state] (q0) {$(\location_0,(0,2,0))$};
      \node[state] (q1) [right of=q0, node distance=3.5cm] {$(\location_1,(0,2,0))$};
      \node[state] (q2) [right of=q1, node distance=4cm] {$(\location_1,\bot)$};

      \node[state] (f1) [below of=q1, node distance=1.4cm] {$c_1 = (f_1,(2,2,0))$};
      \node[state] (f2) [below of=f1, node distance=1.4cm] {$c_3 = (f_1,(1,2,0))$};
      \node[state] (f3) [below of=f2, node distance=1.4cm] {$\phantom{c_5 = }\;(f_1,(0,2,0))$};

      \node[state] (ff1) [right of=f1, node distance=4.5cm] {$(f_2,(2,2,0)) = c_2$};
      \node[state] (ff2) [right of=f2, node distance=4.5cm] {$(f_2,(1,2,0)) = c_4$};
      \node[state] (ff3) [right of=f3, node distance=4.5cm] {$(f_2,(1,2,0)) = c_5$};

      \draw (q0) edge node {\footnotesize $t_0$} (q1);
      \draw (q1) edge node {\footnotesize $t_1$} (q2);

      \draw (q1) edge node {\footnotesize $\rho_1$ ``\textsf{push} $(f_1,(2,2,0))$''} (f1);
      \draw (f1) edge node {\footnotesize $\rho_2$ ``\textsf{push} $(f_1,(1,2,0))$''} (f2);
      \draw (f2) edge node {\footnotesize $\rho_3$ ``\textsf{push} $(f_1,(0,2,0))$''} (f3);

      \draw (f1) edge node {\footnotesize $t_5$} (ff1);
      \draw (f2) edge node {\footnotesize $t_5$} (ff2);
      \draw (f3) edge node {\footnotesize $t_4$} (ff3);

      \draw[dashed] (ff2) edge [right] node {\footnotesize $\epsilon$ ``\textsf{pop}''} (ff1);
      \draw[dashed] (ff3) edge [right] node {\footnotesize $\epsilon$ ``\textsf{pop}''} (ff2);
    \end{tikzpicture}
    \caption{\label{fig:evaluation}
      Exemplary Evaluation of a \its{}}
  \end{figure}
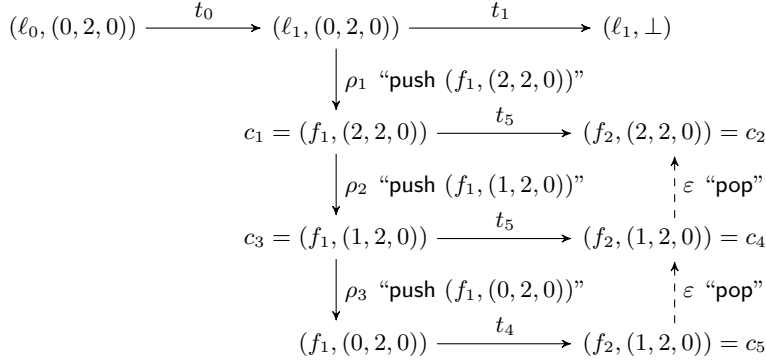
  Reconsider the \its{} from \Cref{fig:ITS} and let us denote states $\valuation\in\Valuation$ as tuples $(\valuation(a),\valuation(x),\valuation(y))\in\ZZ^3$.
  The tree in \Cref{fig:evaluation} shows an evaluation starting in $\tree_{(0,2,0)}$.
  Here, a dashed arrow indicates that a state which was reached via a function call was used to replace $\bot$ via an $\epsilon$-evaluation step.
  Note that these dashed arrows are not part of the actual evaluation tree but they are just depicted to illustrate the construction of the tree.\footnote{This is crucial, because the edges of the evaluation trees correspond to those steps were our \emph{$\rho$-ranking functions} must be (strictly or weakly) decreasing.
    So they have to decrease for steps with transitions like $t_1$ or $t_5$ (where variable values are changed according to the result of a function call), but not for $\epsilon$- (or ``\textsf{pop}'')-steps.
    This is the reason for defining the evaluation of \its{s} via evaluation trees instead of stacks.}
  So for example, if $\valuation_i$ always denotes the state of the configuration $c_i$, then for the configuration $c_2 = (f_2, \valuation_2)$, the value $\valuation_2(a) = 2$ is obtained from $t_5$'s update $\update(a) = a \cdot f_1(\recupdate_2)$ and the value of the return variable in the configuration $c_4$, i.e., $\eval{\update(a)\,[ f_1(\recupdate_1)/\valuation_4(a)]}{\valuation_1} = \eval{a \cdot \valuation_4(a)}{\valuation_1} = \valuation_1(a) \cdot \valuation_4(a) = 2 \cdot 1 = 2$.

  In the next evaluation step, $(\location_1,\bot)$ can be instantiated by considering the return variable in the configuration $c_2$.
  Then, $\bot$ would be replaced by $(0,1,2)$.
  Note that while in this tree, every node has at most one child connected by a $\rho$-edge, in general a node can have several outgoing $\rho$-edges if there exist transitions whose updates contain several function calls (e.g., the naive implementation of the Fibonacci numbers).
\end{example}

The goal of complexity analysis is to derive an upper bound on the number of $t$-evaluation steps starting in the initial tree $\tree_{\initial}$ with the single node $(\location_0,\initial)$.
For any tree $\tree$ and set of transitions $\TSet$, $|\tree|_\TSet$ is the number of edges which are marked by a transition from $\TSet$.
The \emph{runtime complexity} measures how many transitions are evaluated in the worst case in any evaluation tree that results from $\tree_{\initial}$.
In the following, let $\NNC = \NN\cup\set{\omega}$.

\begin{definition}[Runtime Complexity]
  The \emph{runtime complexity} is $\rc:\Valuation\to\NNC$ and $\rc(\initial) = \sup\set{|\tree|_\TSet\mid \tree_{\initial} \step^* \tree}.$
\end{definition}
Here, we count only $t$-edges labeled by transitions (and no $\rho$-edges labeled by function calls) to obtain a natural extension of our previous approach without function calls.
Note that an upper bound on the number of $\rho$-edges can be easily derived from an upper bound on the number of $t$-evaluation steps by multiplying with the branching factor (i.e., by the maximal number of function calls occurring in the update of any transition).
\subsection{Runtime and Size Bounds for \its{s}}
\label{Runtime and Size Bounds}

Now we define our notion of \emph{bounds}.
We only consider bounds which correspond to functions $f$ that are weakly monotonically increasing in all arguments, i.e., where $x \leq y$ implies $f(\ldots x \ldots) \leq f(\ldots y \ldots)$.
In this way, if $f$ and $g$ are both upper bounds, then $f\circ g$ is also an upper bound, i.e., bounds can be ``composed'' easily.
For example, we used this in the introductory \Cref{exa:program} where we inserted the size bound ``$\text{size}(y)$'' into the runtime bound $\log_2(y) + 2$ of the second loop.

In principle, every weakly monotonically increasing function could be used as a bound in our framework.
However, here we restrict ourselves to bounds which are easy to represent and to compute with, and which cover the most prominent complexity classes.
As in \cite{lommen2024TargetingCompletenessComplexity}, bounds can also be \emph{logarithmic}.
In contrast to our earlier papers, we also consider exponential bounds with non-constant bases to represent bounds like $x^x$.
\begin{definition}[Bounds]
  \label{def:bounds}
  The set of \emph{bounds} $\BoundSet$ is the smallest set with $\NNC \subseteq \BoundSet$, $\VSet \subseteq \BoundSet$, and $\{b_1+b_2, \max(b_1,b_2), b_1 \cdot b_2, p^{b_1}, \log_k(b_1)\} \subseteq \BoundSet$ for all $b_1,b_2 \in \BoundSet$, all polynomials $p\in\NN[\VSet]$, and all $k \in \RR_{> 1}$.\footnote{More precisely, instead of $\log_k(b_1)$ we use the function $\lceil\log_k(\max\set{1,b_1})\rceil$ to ensure that bounds are well defined, weakly monotonically increasing, and evaluate to natural numbers.\label{rounding}
  }
\end{definition}

A \emph{runtime bound} $\glo(t)$ over-approximates the number of $t$-evaluations that can occur in an arbitrary evaluation starting in a state $\initial\in\Valuation$, i.e., it is a bound on the number of $t$-edges in any evaluation tree resulting from $\tree_{\initial}$.
In the following, let $\abs{\valuation}$ denote the state with $\abs{\valuation}(v) = \abs{\valuation(v)}$ for all $v\in\VSet$.
\begin{definition}[Runtime Bound]
  \label{def:Runtime Bound}
  $\glo: \TSet\to\BoundSet$ is a \emph{runtime bound} if for all $\initial\in\Valuation$, all $t\in\TSet$, and all trees $\tree$ with $\tree_{\initial} \step^* \tree$, we have $|\tree|_{\set{t}} \leq \eval{\glo(t)}{\abs{\initial}}$.
\end{definition}

\Cref{cor:overapprox_rc} shows that to obtain an upper bound on the runtime complexity, one can compute runtime bounds for each transition separately and add them.

\begin{corollary}[Over-Approximating $\rc$]
  \label{cor:overapprox_rc}
  Let $\glo$ be a runtime bound.
  Then for all states $\initial \in \Valuation$, we have $\rc(\initial) \leq \eval{\sum\nolimits_{t \in \TSet} \glo(t)}{\abs{\initial}}$.
\end{corollary}
\begin{example}
  \label{exa:globalRB}
  In \Cref{fig:ITS}, the transitions $t_0$ and $t_2$ executed at most once, i.e., $\glo(t_0) = \glo(t_2) = 1$.
  In \Cref{Ex:Runtime Bounds}, we will infer a runtime bound with $\glo(t_1) =\glo(t_4) = x$, $\glo(t_3) = \log_2(y + x\cdot x^x) + 2$, and $\glo(t_5) = x^2$.
  This results in a quadratic bound on the runtime complexity of the \its{}.
\end{example}

Our approach performs a \emph{modular} analysis, i.e., parts of the program are analyzed as standalone programs and the results are then lifted to contribute to the overall analysis.
So to compute a runtime bound for a transition $t$, our approach considers all transitions and function calls $\vartheta\in\TSet\cup\FSet$ that can occur before $t$ in evaluations, and it needs \emph{size bounds} $\Size(\vartheta,v)$ to over-approximate the absolute values that the variables $v \in \VSet$ may have \emph{after} these ``previous'' transitions and function calls $\vartheta$.
We call $\RVSet = (\TSet\cup\FSet) \times \VSet$ the set of \emph{result variables}.
Note that in contrast to runtime bounds (and to our earlier papers), we now also have to capture the effect of function calls $\FSet$ via size bounds.
\begin{definition}[Size Bound]
  A function $\Size:\RVSet\to\BoundSet$ is called a \emph{size bound} if for all $(\vartheta,v)\in\RVSet$, all states $\initial\in\Valuation$, and all trees $\tree$ with $\tree_{\initial} \step^* \tree$ containing a path $(\location_0,\initial) \to \dots \to_\vartheta(\_,\valuation) \text{ with } \valuation\neq\bot$, we have $\abs{\valuation}(v) \leq \eval{\Size(\vartheta,v)}{\abs{\initial}}$.
\end{definition}

\begin{example}
  In \Cref{fig:ITS}, $\Size(t_0,x) = x$ is a size bound, since the value of $x$ after evaluating $t_0$ is bounded by the initial value of $x$.
  (\Cref{Ex:SizeBounds} will show how to com\-pute such size bounds.)
  Similarly, we have $\Size(t_2,x) = 1$ and $\Size(t_2,y) = y + x\cdot x^x$, see \Cref{exa:NonTrivialSB}.
  The size bound $\Size(\rho_1,a) = x$ (see \Cref{Ex:SizeBounds}) expresses that the value of $a$ after executing the function call $\rho_1$ is bounded by the initial value of $x$.
\end{example}

\section{Modular Computation of Runtime Bounds}
\label{sect:RB}
Now we introduce our modular approach for the computation of runtime bounds.
To be precise, we infer runtime bounds for subprograms $\TSet'$ and then lift them to runtime bounds for the full program.
For any non-empty $\TSet' \subseteq \TSet$, let $\LSet_{\TSet'} = \{\location\in\LSet\mid(\location,\wildcard,\wildcard,\wildcard\,)\in\TSet'\}$ contain all start locations of transitions from $\TSet'$.

Recall that a global runtime bound $\glo(t)$ over-approximates how many times the transition $t\in\TSet$ may be evaluated when starting an evaluation of $\TSet$ from the initial location $\location_0\in\LSet$.
To make this explicit, instead of $\glo(t)$ we could also write $\glo^{t,\TSet}(\location_0)$.
Now we are also interested in \emph{local runtime bounds}, where one takes a subprogram $\TSet'\subseteq\TSet$ and a location $\ell$ from $\LSet_{\TSet'}$ into account.
Then $\glo^{t,\TSet'}(\location)$ over-approximates the number of applications of the transition $t\in\TSet'$ in any run of $\TSet'$ starting in the location $\location\in\LSet_{\TSet'}$.
To highlight that these runtime bounds may be ``local'' (i.e., that they can regard arbitrary subprograms $\TSet'$), we add the\linebreak
subscript ``loc''.
So a \emph{local runtime bound} is a function $\locParameter{t}{\TSet'}:\LSet_{\TSet'}\to\BoundSet$.
If $\TSet' =\linebreak
  \TSet$, then $\locParameter{t}{\TSet}(\location_0)$ corresponds to a global runtime bound $\glo(t)$ of $t$.
We define local runtime bounds as functions from locations to bounds (rather than from transitions, as with global runtime bounds) in order to simplify the presentation later on when inferring local runtime bounds from ranking functions, because ranking functions are also functions from locations to bounds, see \Cref{thm:localRF}.

So for any state $\initial\in\Valuation$, $\eval{\locParameter{t}{\TSet'}(\location)}{\abs{\initial}}$ is an upper bound on the number of $t$-edges that can occur in any evaluation tree resulting from the initial single-node tree $(\location, \initial)$, if only transitions from $\TSet'$ and $\epsilon$-steps are used, i.e., one only executes the subprogram $\TSet'$.
However, local runtime bounds do not consider how often such a local evaluation of the subprogram $\TSet'$ is started or how large the variables are before starting such a local evaluation.
If $t$ and $\TSet'$ are clear from the context, we just write $\loc$ instead of $\locParameter{t}{\TSet'}$.
\begin{definition}[Local Runtime Bound]
  \label{def:Local Runtime Bound}
  Let $\emptyset \neq \TSet' \subseteq \TSet$ be a set of transitions and let $t\in\TSet'$.
  Then $\loc:\LSet_{\TSet'}\to\BoundSet$ is a \emph{local runtime bound} for $t$ w.r.t.\ $\TSet'$ if for all $\initial\in\Valuation$, all $\location\in\LSet_{\TSet'}$, and all trees $\tree$ with $(\set{(\location,\initial)},\emptyset) \step_{\TSet'\cup\set{\epsilon}}^* \tree$, we have $\abs{\tree}_{\set{t}} \leq \eval{\loc(\location)}{\abs{\initial}}$.
  To make this explicit, we also write $\locParameter{t}{\TSet'}$.
\end{definition}
For readability, in contrast to our previous work \cite{lommen2022AutomaticComplexityAnalysis,lommen2023TargetingCompletenessUsing,lommen2024TargetingCompletenessComplexity}, \Cref{def:Local Runtime Bound} considers arbitrary initial states $\initial\in\Valuation$, but it could also be refined to only consider states $\initial\in\Valuation$ where $(\location,\initial)$ is reachable in the full program with all transitions $\TSet$.
Note that the actual evaluation relation is considered both for local runtime bounds and also for local size bounds in \Cref{sect:SB}, i.e., here of course one also takes the guards of transitions into account.

In \Cref{sec:Ranking Functions} we introduce a novel form of $\rho$-\emph{ranking functions} to derive local runtime bounds for \its{s} with function calls automatically.
Then, we show in \Cref{sec:lifting} how global runtime bounds can be inferred from local runtime bounds.

\subsection{Ranking Functions}
\label{sec:Ranking Functions}
Ranking functions are widely used to analyze termination or runtime complexity of programs (e.g., \cite{ben-amram2017MultiphaseLinearRankingFunctions,ben-amram2019MultiphaseLinearRankingFunctions,heizmannRankingTemplatesLinear2015,giesl2022ImprovingAutomaticComplexity,bradleyPolyrankingPrinciple2005,podelskiCompleteMethodSynthesis2004}).
The idea of ranking functions is to construct a well-founded relation by mapping program configurations to numbers.
Typically, SMT solvers are used to find such a suitable mapping automatically.
We first consider programs without \emph{recursive} function calls and recapitulate classical\linebreak
ranking functions in our setting of \its{s}.
Afterwards, we extend ranking functions to a novel form of \emph{function call ranking functions}, which allow us to derive local runtime bounds for \its{s} with arbitrary (possibly recursive) function calls.

Recall that $\fun(t)$ denotes the set of function calls in the transition $t$ and $\trans(\rho)$ is the set of all transitions with the function call $\rho$.
For $\TSet' \subseteq \TSet$, let $\fun(\TSet') = \bigcup_{t \in \TSet'} \fun(t)$ denote the set of all function calls $\rho\in\fun(t)$ for transitions $t\in\TSet'$.
Moreover, $\fun(\LSet_{\TSet'})$ is the set of all function calls $\location(\,\wildcard \,)$ with $\location\in \LSet_{\TSet'}$,\linebreak
i.e., all function calls that lead into the subprogram $\TSet'$.
We also refer to $\fun(\TSet')\cap\fun(\LSet_{\TSet'})$ as the set of \emph{recursive}
function calls of the subprogram $\TSet'$ and to $\TSet' \cap \trans(\fun(\LSet_{\TSet'}))$ as the set of \emph{recursive} transitions of the subprogram $\TSet'$.

\paragraph{Classical Ranking Function:}
Before we introduce our novel class of ranking functions for \its{s}, we recapitulate classical ranking functions and adapt them to the setting of \its{s}.
As mentioned, we first consider subprograms $\TSet'$ which do not contain recursive function calls for jumps ``within'' $\TSet'$, i.e., where $\fun(\TSet') \cap \fun(\LSet_{\TSet'}) = \emptyset$.
A function $r_{\mathrm{d}}: \LSet\to\ZZ[\VSet]$ is a classical \emph{ranking function} for the \emph{\underline{\textbf{d}}ecreasing} transition $t$ w.r.t.\ the subprogram $\TSet'$, if every evaluation step with $t$ decreases the value of $r_{\mathrm{d}}$, where $r_{\mathrm{d}}$'s value is positive, and evaluation steps with the (other) transitions of $\TSet'$ do not increase the value of $r_{\mathrm{d}}$.
Then for any $\TSet'$-evaluation tree, $r_{\mathrm{d}}(\location)$ is a bound on the number of edges labeled with the decreasing transition $t$.

We also use the relations $\to_t$ and $\to_\rho$ without referring to an actual evaluation tree.
Thus, we say that $(\location,\valuation) \to_\vartheta (\location',\valuation')$ holds for some $\vartheta \in \TSet \cup \FSet$ if there exists an evaluation $\tree_{\initial}\step_{\TSet\cup\set{\epsilon}}^* \tree$ for some initial state $\initial\in\Valuation$ such that $\tree$ contains an edge $(\location,\valuation) \to_\vartheta (\location',\valuation')$.
In the following definition, we extend the evaluation of an arithmetic expression $e$ to the ``undefined'' state $\bot$ by defining $\eval{e}{\bot} = 0$.
\begin{definition}[Ranking Function]
  \label{def:Ranking Function}
  Let $\emptyset \neq \TSet'\!\subseteq \TSet\!$ with $\fun(\TSet') \,\cap\, \fun(\LSet_{\TSet'})\linebreak
    = \emptyset$, and let $t\in\TSet'$.
  Then $r_{\mathrm{d}}: \LSet\to\ZZ[\VSet]$ is a \emph{ranking function} (RF) for the transition $t$ w.r.t.\ $\TSet'$ if for all evaluation steps $(\location,\valuation)\to_{t'}(\location',\valuation')$, we have: \TabPositions{2.6cm}
  \begin{enumerate}[leftmargin=*]
    \item if $t'\in\TSet'$, then \tab $\eval{r_{\mathrm{d}}(\location)}{\valuation} \geq \eval{r_{\mathrm{d}}(\location')}{\valuation'}$ \label{rf:a}
    \item if $t' = t$, then \tab $\eval{r_{\mathrm{d}}(\location)}{\valuation} > \eval{r_{\mathrm{d}}(\location')}{\valuation'}$ and $\eval{r_{\mathrm{d}}(\location)}{\valuation} > 0$ \label{rf:b}
  \end{enumerate}
\end{definition}

So \Cref{def:Ranking Function} does not impose any requirements on how the value of $r_{\mathrm{d}}$ changes when following edges labeled with function calls.
The reason is that due to the requirement $\fun(\TSet') \cap \fun(\LSet_{\TSet'}) = \emptyset$, all function calls of $\TSet'$ lead outside this subprogram and thus, they are irrelevant for local runtime bounds w.r.t.\ $\TSet'$.
Note that if $t'$'s update $\update$ contains function calls, then in general it is not decidable whether $(\location,\valuation)\to_{t'}(\location',\valuation')$ holds.
Thus, to over-approximate $\to_{t'}$ in our automation, we consider a modified update $\update'$ where all function calls are replaced by fresh ``non-deterministic'' values.
To ease the automation of our approach, in practice we restrict ourselves to linear polynomial ranking functions and use the SMT solver \tool{Z3} \cite{moura2008Z3SMTSolver} to infer ranking functions automatically.
Our modular approach allows us to consider program parts separately, such that using linear ranking functions usually suffices even if the overall program has non-linear runtime.
Note that when lifting local to global runtime bounds (in \Cref{sec:lifting}), we use size bounds that indeed take the results of ``previous'' function calls into account (see \Cref{sect:SB} for the computation of size bounds).

\begin{example}
  \label{exa:rf}
  We compute a (classical) ranking function for the transition $t_1$ w.r.t.\ the subprogram $\TSet' = \set{t_1}$ from \Cref{fig:ITS}:
  We have $\fun(\TSet') = \{\rho_1\}$, but the location $f_1$ of $\rho_1$ is not in $\LSet_{\TSet'} = \{ \ell_1 \}$.
  Thus, $\fun(\TSet') \cap \fun(\LSet_{\TSet'}) = \{\rho_1\} \cap \emptyset = \emptyset$, i.e., $\TSet'$ does not have any recursive function calls, but all function calls of $\TSet'$ \emph{leave} the subprogram $\TSet'$.
  An RF for $t_1$ w.r.t.\ $\TSet' = \set{t_1}$ is $r_{\mathrm{d}}(\location_1) = x$.
  The RF can always map all remaining locations outside the subprogram $\TSet'$ to $0$.
\end{example}

\Cref{thm:localRF}
shows how to obtain a local runtime bound from a (classical) ranking function.
We use $\approx{\cdot}$ to transform a polynomial into a (weakly monotonically increasing) bound from $\BoundSet$ by taking the absolute values of the coefficients, e.g., for the polynomial $x - y$ we obtain the bound $\approx{x-y} = x + y$.
\begin{restatable}[Local Runtime Bounds by RFs]{lemma}{thmLocalRF}
  \label{thm:localRF}
  Let $\emptyset \neq \TSet' \subseteq \TSet$ where $\fun(\TSet') \cap \fun(\LSet_{\TSet'}) = \emptyset$, and let $t\in\TSet'$.
  Moreover, let $r_{\mathrm{d}}$ be an RF for $t$ w.r.t.\ $\TSet'$.
  Then $\loc:\LSet_{\TSet'}\to\BoundSet$ is local runtime bound for $t$ w.r.t.\ $\TSet'$, where for all $\location\in\LSet_{\TSet'}$, we define $\loc(\location) = \approx{r_{\mathrm{d}}(\location)}$.
\end{restatable}
\makeproof{thm:localRF}{
  \thmLocalRF*
  \begin{myproof}
    Let $(\set{(\location,\valuation)},\emptyset) \step_{\TSet'\cup\set{\epsilon}}^* \tree$ be an evaluation in the subprogram $\TSet'$ with $\location\in\LSet_{\TSet'}$ and an arbitrary state $\valuation\in\Valuation$.
    We have to prove that
    \begin{equation}
      \abs{\tree}_{\{t\}} \leq \eval{r_{\mathrm{d}}(\location)}{\abs{\valuation}} = \eval{\locParameter{t}{\TSet'}(\location)}{\abs{\valuation}}
    \end{equation}
    holds.
    To this end, we consider an arbitrary evaluation $(\set{(\location,\valuation)},\emptyset) \step_{\TSet'\cup\set{\epsilon}}^* \tree$.
    Since we have $\fun(\TSet') \cap \fun(\LSet_{\TSet'}) = \emptyset$, the transition $t$ can only occur at most $\eval{r_{\mathrm{d}}(\location)}{\abs{\valuation}}$ times by Requirements $\Cref{rf:a}$ and $\Cref{rf:b}$ in \Cref{def:Ranking Function}.
    To be precise, $\Cref{rf:a}$ and $\Cref{rf:b}$ yield a well-founded relation where each chain has length at most $\eval{r_{\mathrm{d}}(\location)}{\abs{\valuation}}$.

  \end{myproof}
}

\begin{example}
  \label{exa:localRBRF}
  With the ranking function of \Cref{exa:rf}, \Cref{thm:localRF} yields the local runtime bound $\locParameter{t_1}{\set{t_1}}\,(\location_1) = \approx{r_{\mathrm{d}}(\location_1)} = x$.
\end{example}

\paragraph{Function Call Ranking Functions:}
Now we consider arbitrary \its{s} with possibly recursive function calls.
To obtain local runtime bounds for such \its{s}, we introduce the novel concept of ``\emph{function call ranking functions}'' (\rrf{s}).
In contrast to \emph{classical} ranking functions, \rrf{s} also contain an explicit bound on the number of \emph{function} calls.
More precisely, a \rrf{} $\langle r_{\mathrm{d}},r_{\mathrm{tf}},r_{\mathrm{f}} \rangle$ combines a classical ranking function $r_{\mathrm{d}}$ with two additional ranking functions $r_{\mathrm{tf}}: \LSet\to\ZZ[\VSet]$ and $r_{\mathrm{f}}: \LSet\to\ZZ[\VSet]$.
While $r_{\mathrm{d}}$ yields a bound for a \underline{\textbf{d}}ecreasing transition $t$ w.r.t.\ $\TSet'$, $r_{\mathrm{tf}}$ yields a bound for the \underline{\textbf{t}}ransitions from $\TSet'$ with recursive \underline{\textbf{f}}unction calls and $r_{\mathrm{f}}$ yields a bound on these recursive \underline{\textbf{f}}unction calls themselves.
We distinguish these three ranking functions because they have different influences on the overall bound for the possible number of applications of $t$ in the subprogram $\TSet'$.
For example, $r_{\mathrm{f}}$ has an exponential influence on the runtime while the other two ranking functions only have a polynomial impact, see \Cref{thm:localRRF}.

More precisely, for a set of function calls $\FSet'$, let $\trans(\FSet') = \bigcup_{\rho \in \FSet'} \trans(\rho)$ be the set of all transitions $t \in \trans(\rho)$ for function calls $\rho \in \FSet'$.
Recall that $\fun(\LSet_{\TSet'})$ is the set of all function calls $\ell(\wildcard)$ with a location $\ell \in \LSet_{\TSet'}$.
Thus, $\trans(\fun(\LSet_{\TSet'}))$ is the set of all transitions which contain a function call $\location(\wildcard)$ with $\location \in \LSet_{\TSet'}$.
Then for any location $\location \in \LSet$ and any $\TSet'$-evaluation tree, $r_{\mathrm{tf}}(\location)$ yields a bound on the number of edges labeled with transitions from $\TSet' \cap \trans(\fun(\LSet_{\TSet'}))$ on paths starting with $(\location,\wildcard\,)$, where these paths may contain both steps with transitions and steps with function calls.
So $\TSet' \cap \trans(\fun(\LSet_{\TSet'}))$ are recursive \underline{\textbf{t}}ransitions from $\TSet'$ whose \underline{\textbf{f}}unction call jumps into $\TSet'$ again.

Furthermore, for any location $\location \in \LSet$ and any $\TSet'$-evaluation tree, $r_{\mathrm{f}}(\location)$ is a bound on the number of $\rho$-edges with function calls that jump into $\TSet'$ again on paths starting with $(\location,\wildcard\,)$.
So there are recursive function calls from $\fun(\TSet') \cap \fun(\LSet_{\TSet'})$.
In \Cref{exa:evaluation}, these function calls correspond to vertical edges, whereas steps with transitions correspond to horizontal edges.

As for classical RFs, \Cref{def:Recursive Ranking Function}\Cref{rrf:a} requires that $r_{\mathrm{d}}$, $r_{\mathrm{tf}}$, and $r_{\mathrm{f}}$ do not increase for\linebreak
edges of a $\TSet'$-evaluation tree.
Furthermore, to ensure the previously mentioned\linebreak
properties, as for classical RFs, \Cref{rrf:b} requires that $r_{\mathrm{d}}$ is decreasing and positive for\linebreak
edges labeled with $t$ (if the decreasing transition $t$ is not recursive, i.e., it does not contain any recursive function calls).
Condition \Cref{rrf:c} imposes these requirements for $r_{\mathrm{tf}}$ and edges labeled with recursive transitions from $\TSet' \cap \trans(\fun(\LSet_{\TSet'}))$.
In particular, this is also required for $t$ if $t$ is a recursive transition and thus, it was not already handled in \Cref{rrf:b}.
Finally, \Cref{rrf:d} requires these properties for $r_{\mathrm{f}}$ and edges labeled with recursive function calls from $\fun(\TSet') \cap \fun(\LSet_{\TSet'})$.
\begin{definition}[Function Call Ranking Function]
  \label{def:Recursive Ranking Function}
  Let $\emptyset \neq \TSet' \subseteq \TSet$ and let $t\in\TSet'$.
  Then $\langle r_{\mathrm{d}},r_{\mathrm{tf}},r_{\mathrm{f}} \rangle$ with $r_{\mathrm{d}},r_{\mathrm{tf}},r_{\mathrm{f}}: \LSet\to\ZZ[\VSet]$ is a \emph{function call ranking function} (\rrf{}) for $t$ w.r.t.\ $\TSet'$ if for all evaluation steps $(\location,\valuation)\to_\vartheta(\location',\valuation')$ with $\vartheta\in\TSet'$ or $\vartheta\in\fun(\TSet') \cap \fun(\LSet_{\TSet'})$, we have: \TabPositions{5.2cm,8.5cm}
  \begin{enumerate}[leftmargin=*]
    \item for all $i\in\set{\mathrm{d},\mathrm{tf},\mathrm{f}}$ \tab $\eval{r_i(\location)}{\valuation} \,\geq \eval{r_i(\location')}{\valuation'}$ \tab \label{rrf:a}
          \vspace{0.1cm}
    \item if $\vartheta = t$ and $t\not\in \trans(\fun(\LSet_{\TSet'}))$, then \label{rrf:b}
          \newline
          \vspace{0.1cm} \tab $\eval{r_{\mathrm{d}}(\location)}{\valuation} > \eval{r_{\mathrm{d}}(\location')}{\valuation'}$ \tab and\; $\eval{r_{\mathrm{d}}(\location)}{\valuation} > 0$ \vspace{0.1cm}
    \item if $\vartheta\in \TSet' \cap \trans(\fun(\LSet_{\TSet'}))$, then \tab $\eval{r_{\mathrm{tf}}(\location)}{\valuation} \!> \eval{r_{\mathrm{tf}}(\location')}{\valuation'}$ \tab and \label{rrf:c}
          $\eval{r_{\mathrm{tf}}(\location)}{\valuation} \!> 0$\vspace{0.1cm}
    \item if $\vartheta\in\fun(\TSet') \cap \fun(\LSet_{\TSet'})$, then \tab $\eval{r_{\mathrm{f}}(\location)}{\valuation} \,> \eval{r_{\mathrm{f}}(\location')}{\valuation'}$ \tab and\; $\eval{r_{\mathrm{f}}(\location)}{\valuation} \,> 0$ \label{rrf:d}
  \end{enumerate}
\end{definition}
Note that $r_{\mathrm{d}}$ can be set to $0$ for all locations if $t\in \trans(\fun(\LSet_{\TSet'}))$.

\begin{example}
  \label{exa:rrf}
  We now compute \rrf{s} for $t_1$ w.r.t.\ $\set{t_1}$ and for $t_5$ w.r.t.\ $\set{t_4,t_5}$ in the program from \Cref{fig:ITS}.
  \begin{itemize}
    \item First, we consider $\TSet' = \set{t_1}$.
          As in \Cref{exa:rf}, we have $\fun(\TSet') \cap \fun(\LSet_{\TSet'}) = \emptyset$ and $\TSet' \, \cap \, \trans(\fun(\LSet_{\TSet'})) = \emptyset$, since there is no function call with the location $\LSet_{\TSet'} = \{ \location_1 \}$, i.e., $\fun(\LSet_{\TSet'}) = \emptyset$.
          A \rrf{} for $t_1$ is $r_{\mathrm{tf}}(\location_1) = r_{\mathrm{f}}(\location_1) = 0$ and $r_{\mathrm{d}}(\location_1) = x$.
          Again, locations outside $\TSet'$ can always be mapped to $0$.
          So, here the \rrf{} corresponds to the classical ranking function from \Cref{exa:rf}.
    \item For the subprogram $\TSet' = \set{t_4,t_5}$ we have $\fun(\TSet') \cap \fun(\LSet_{\TSet'}) = \{ \rho_2 \}$ and $\TSet' \cap \trans(\fun(\LSet_{\TSet'})) = \{ t_5 \}$.
          Thus, a \rrf{} for $t_5$ is $r_{\mathrm{d}}(f_1) = r_{\mathrm{d}}(f_2) = 0$ (as $t_5\in \trans(\fun(\LSet_{\TSet'}))$), $r_{\mathrm{tf}}(f_1) = 1$ and $r_{\mathrm{tf}}(f_2) = 0$ (since $t_5$ was not already handled by $r_{\mathrm{d}}$, $t_5$ must be decreasing for $r_{\mathrm{tf}}$), and $r_{\mathrm{f}}(f_1) = a$, $r_{\mathrm{f}}(f_2) = 0$ (to make $\rho_2$ decreasing, since $(f_1,\valuation) \to_{\rho_2} (f_1, \valuation')$ implies $\eval{r_{\mathrm{f}}(f_1)}{\valuation} = \valuation(a) > \valuation(a-1) = \valuation'(a) = \eval{r_{\mathrm{f}}(f_1)}{\valuation'}$).
  \end{itemize}
\end{example}

Now we show how to construct a local runtime bound from a \rrf{}.
To simplify the presentation, here we restrict ourselves to transitions which have at most one function call in their update and refer to\report{ App.\ B}\paper{ \cite[App.\ B]{report}} for the general case which handles transitions with arbitrary many function calls.

For a local runtime bound, we have to over-approximate how many edges labeled with $t$ can occur in a $\TSet'$-evaluation tree starting with a configuration of the form $(\location,\wildcard\,)$.
As mentioned, the ranking functions $r_{\mathrm{d}}$, $r_{\mathrm{tf}}$, and $r_{\mathrm{f}}$ influence the local runtime bound in different ways:
If every path has at most $n_0$ edges labeled with the transition $t$ (if $t$ is not a recursive transition from $\trans(\fun(\LSet_{\TSet'}))$), $n_1$ edges labeled with the recursive transitions from $\TSet' \cap \trans(\fun(\LSet_{\TSet'}))$, and $n_2$ edges labeled with recursive function calls, then $\mathcal{R}_{n_0}(n_1,n_2)$ over-approximates the number of $t$-edges in any $\TSet'$-evaluation tree, where $\mathcal{R}_{n_0}(n_1,n_2)$ is defined via the following recurrence:

\vspace*{-.3cm}

\[\mbox{\small
    $\mathcal{R}_{n_0}(n_1,n_2) = \left\{
      \begin{array}{ll@{\hspace*{0.5cm}}r}
         & n_0,                                                                       & \text{if $n_1 = 0$ or $n_2 = 0$} \\
         & 1 + n_0 + \mathcal{R}_{n_0}(n_1 - 1,n_2) + \mathcal{R}_{n_0}(n_1,n_2 - 1), & \text{otherwise}                 \\
      \end{array}
      \right.$}
\]
\begin{figure}[t]
  \center
  \begin{tikzpicture}[->,>=stealth',shorten >=1pt,auto]
    \node[state, minimum size=0.5cm, node distance=4.5cm] (qInit) {$N_0$};
    \node[state,draw=none] (h1) [right of=qInit, node distance=1.5cm] {$\cdots$};
    \node[state, label=above:{\footnotesize $\mathcal{R}_{n_0}(n_1,n_2)$}, minimum size=0.5cm,right of=h1, node distance=1.5cm] (q0) {$N_1$};
    \node[state, minimum size=0.5cm, node distance=4.5cm] (q2) [right of=q0] {$N_2$};
    \node[state,draw=none] (q1) [right of=q2, node distance=1.5cm] {$\cdots$};
    \node[state, label=above:{\footnotesize $\mathcal{R}_{n_0}(n_1 - 1,n_2)$}] (q2a) [right of=q1, minimum size=0.5cm, node distance=1.5cm] {$N_3$};

    \node[state,label=below:{\footnotesize $\mathcal{R}_{n_0}(n_1,n_2 - 1)$}, minimum size=0.5cm] (q5) [below of=q0,node distance=1.5cm] {};

    \draw (qInit) edge [below] node {\scriptsize $t$} (h1);
    \draw (h1) edge [below] node {\scriptsize $t$} (q0);
    \draw (q2) edge [below] node {\scriptsize $t$} (q1);
    \draw (q1) edge [below] node {\scriptsize $t$} (q2a);
    \draw (q0) edge [below] node {\scriptsize $\TSet' \cap \trans(\fun(\LSet_{\TSet'}))$} (q2);

    \draw (q0) edge node [align=center,left] {\scriptsize $\rho$} (q5);

    \tikzset{
      position label/.style={
          below = 3pt, text height = 1.5ex, text depth = 1ex
        }, brace/.style={
          decoration={brace, mirror}, decorate
        }
    }

    \node[state,draw=none,node distance=0.25cm] (qh1) [below right of=q2] {};
    \node[state,draw=none,node distance=0.25cm] (qh2) [below left of=q2a] {};
    \node[state,draw=none,node distance=0.25cm] (qh3) [below right of=qInit] {};
    \node[state,draw=none,node distance=0.25cm] (qh4) [below left of=q0] {};

    \draw [brace,-,shorten >=0pt] (qh1.south east) -- (qh2.south west) node [position label, pos=0.5] {$d_2$ steps};
    \draw [brace,-,shorten >=0pt] (qh3.south east) -- (qh4.south west) node [position label, pos=0.5] {$d_1$ steps};

  \end{tikzpicture}
  \caption{Illustration of $\mathcal{R}_{n_0}(n_1,n_2)$}
  \label{fig:illustration_recurrence}
\end{figure}
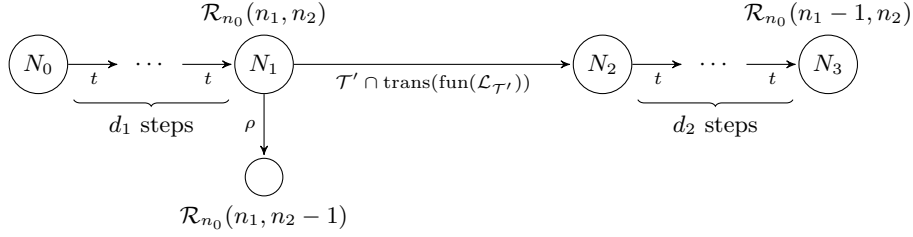

\vspace*{-.2cm}

\noindent
This is shown by induction on $n_1 + n_2$:
If $n_1 = 0$ or $n_2 = 0$, then there is no recursive function call and thus, there can be at most $n_0$ edges labeled with the decreasing transition $t$.
The induction step is illustrated in \Cref{fig:illustration_recurrence}.
First consider the case where $t$ is not a recursive transition.
Then the path from the root to the first node $N_1$ where a function is called recursively uses at most $d_1 \leq n_0$ edges labeled with $t$.
Due to our restriction on the number of function calls in the update, $N_1$ has at most one outgoing edge labeled with a recursive function call $\rho$ and one outgoing edge to a node $N_2$ labeled with the corresponding recursive transition from $\TSet' \cap \trans(\fun(\LSet_{\TSet'}))$ whose update contains $\rho$.
The function call $\rho$ leads to a subtree with at most $\mathcal{R}_{n_0}(n_1,n_2 - 1)$ many $t$-edges by the induction hypothesis.
The path from $N_2$ to the next node $N_3$ where a function might be called uses at most $d_2$ edges labeled with $t$, where $d_1 + d_2 \leq n_0$.
Finally, the subtree starting in the node $N_3$ has at most $\mathcal{R}_{n_0}(n_1-1,n_2)$ many $t$-edges by the induction hypothesis.
Thus, the number of $t$-edges in the full tree is at most
\begin{align*}
  \abs{\tree}_{\{t\}} & \leq d_1 + \mathcal{R}_{n_0}(n_1,n_2 - 1) + d_2 + \mathcal{R}_{n_0}(n_1-1,n_2) \\
                      & \leq n_0 + \mathcal{R}_{n_0}(n_1 - 1,n_2) + \mathcal{R}_{n_0}(n_1,n_2 - 1).
\end{align*}

If $t$ is recursive, we have $d_1 = d_2 = n_0 = 0$.
However, the step from $N_1$ to $N_2$ might be done with $t$.
Thus, we obtain $\abs{\tree}_{\{t\}} \leq 1 + \mathcal{R}_{n_0}(n_1,n_2 - 1) + \mathcal{R}_{n_0}(n_1-1,n_2)$.
So in both cases, we have $\abs{\tree}_{\{t\}} \leq 1 + n_0 + \mathcal{R}_{n_0}(n_1 - 1,n_2) + \mathcal{R}_{n_0}(n_1,n_2 - 1)$.

As shown in\report{ App.\ A}\paper{ \cite[App.\ A]{report}}, $n_0 + n_2 \cdot (1 + 2\cdot n_0) \cdot n_1^{n_2}$ is an over-approximating closed form solution of $\mathcal{R}_{n_0}(n_1,n_2)$.
Hence, instantiating this closed form with the ranking functions yields the desired local runtime bound.
As mentioned, a generalized version of this theorem for transitions with arbitrary many function calls in their updates can be found in\report{ App.\ B}\paper{ \cite[App.\ B]{report}}.\footnote{An alternative approach would be to add suitable transitions in order to transform any program into a program with at most one function call per transition.
  When using a \rrf{}
  on the transformed program, \Cref{thm:localRRF} would result in a similar local runtime bound as when using the \rrf{} on the original program and computing the local runtime bound via the generalized version of \Cref{thm:localRRF} in\report{ App.\ B.}\paper{
    \cite[App.\ B]{report}.}}

\begin{restatable}[Local Runtime Bounds by \rrf{}s]{theorem}{thmLocalRRF}
  \label{thm:localRRF}
  Let $\emptyset \neq \TSet' \subseteq \TSet$ such that $| \fun(\update) | \leq 1$ holds for every update $\update$ of the transitions in $\TSet'$.
  Moreover, let $t\in\TSet'$ and $\langle r_{\mathrm{d}},r_{\mathrm{tf}},r_{\mathrm{f}} \rangle$ be a \rrf{} for $t$ w.r.t.\ $\TSet'$.
  Then $\loc:\LSet_{\TSet'}\to\BoundSet$ is local runtime bound for $t$ w.r.t.\ $\TSet'$, where for all $\location\in\LSet_{\TSet'}$, we define $\loc(\location)$ as:
  \[
    \approx{r_{\mathrm{d}}(\location)} \; + \; \approx{r_{\mathrm{f}}(\location)} \; \cdot \; \left(1 +
    2\cdot\approx{r_{\mathrm{d}}(\location)}\right) \; \cdot \;
    \approx{r_{\mathrm{tf}}(\location)}^{\approx{r_{\mathrm{f}}(\location)}}
  \]
\end{restatable}
\makeproof{thm:localRRF}{
  \thmLocalRRF*
  \begin{myproof}
    Let $(\set{(\location,\valuation)},\emptyset) \step_{\TSet'\cup\set{\epsilon}}^* \tree$ be an evaluation in the subprogram $\TSet'$ with $\location\in\LSet_{\TSet'}$ and an arbitrary state $\valuation\in\Valuation$.
    We have to prove that
    \begin{equation}
      \abs{\tree}_{\{t\}} \leq \eval{\locParameter{t}{\TSet'}(\location)}{\abs{\valuation}} \label{eq:goalRRF}
    \end{equation}
    holds.
    To this end, we consider the following recurrence:
    \[\mbox{\small
        $\mathcal{R}_{n_0}(n_1,n_2) = \left\{
          \begin{array}{ll@{\hspace*{0.5cm}}r}
             & n_0,                                                                       & \text{if $n_1 = 0$ or $n_2 = 0$} \\
             & 1 + n_0 + \mathcal{R}_{n_0}(n_1 - 1,n_2) + \mathcal{R}_{n_0}(n_1,n_2 - 1), & \text{otherwise}                 \\
          \end{array}
          \right.$}
    \]
    As shown in \Cref{sect:RB}, by induction on $n_1 + n_2$ one can prove that $\mathcal{R}_{n_0}(n_1,n_2)$ over-approximates the number of $t$-edges in any $\TSet'$-evaluation tree (i.e., $\abs{\tree}_{\{t\}} \leq \mathcal{R}_{n_0}(n_1,n_2)$), provided that every path has at most $n_0$ edges labeled with the transition $t$ and $t\not\in \TSet' \cap \trans(\fun(\LSet_{\TSet'}))$, $n_1$ edges labeled with the recursive transitions from $\TSet' \cap \trans(\fun(\LSet_{\TSet'}))$, and $n_2$ edges labeled with recursive function calls.
    Now we show that $n_0 + n_2 \cdot (1 + 2\cdot n_0) \cdot n_1^{n_2}$ is an over-approximating closed form solution of $\mathcal{R}_{n_0}(n_1,n_2)$.
    Instantiating this closed form with the ranking functions yields the desired local runtime bound.

    Let us abbreviate $f(n_0,n_1,n_2) = n_0 + n_2 \cdot (1 + 2\cdot n_0) \cdot n_1^{n_2}$.
    We show that for all $n_0, n_1, n_2 \in \NN$ we have $\mathcal{R}_{n_0}(n_1,n_2) \leq f(n_0,n_1,n_2)$ by induction on $n_2$.

    If $n_2 = 0$, then we have $\mathcal{R}_{n_0}(n_1,n_2) = f(n_0,n_1,0) = n_0$.
    Otherwise, if $n_1 = 0$ and $n_2 > 0$, then we also have $\mathcal{R}_{n_0}(n_1,n_2) = f(n_0,0,n_2) = n_0$.
    Finally, if $n_1 > 0$ and $n_2 > 0$, then we have

    \vspace*{-.2cm}

    \begin{align}
      \mathcal{R}_{n_0}(n_1,n_2) & = 1 + n_0 + \mathcal{R}_{n_0}(n_1 - 1,n_2) + \mathcal{R}_{n_0}(n_1,n_2 - 1) \nonumber \\
                                 & \leq 1 + n_0 + \mathcal{R}_{n_0}(n_1 - 1,n_2) +
      f(n_0,n_1,n_2 - 1) \tag{by the induction hypothesis} \nonumber \\
                                 & \leq \mathcal{R}_{n_0}(0,n_2) + \sum_{i = 0}^{n_1 - 1} \left( 1 + n_0 +
      f(n_0,n_0 - i,n_2 - 1) \right). \label{sumClaim}
    \end{align}

    \noindent
    The last step \eqref{sumClaim} is clear if $n_1 = 1$.
    Otherwise, if $n_1 > 1$, the reason for \eqref{sumClaim} is that we have

    \vspace*{-.5cm}

    \begin{align}
      \phantom{\mathcal{R}_{n_0}(n_1,n_2)} &
      \phantom{\leq}\;\; 1 + n_0 + f(n_0,n_1,n_2 - 1) + \mathcal{R}_{n_0}(n_1 - 1,n_2) \nonumber \\
                                           & = 1 + n_0 + f(n_0,n_1,n_2 - 1)\; + \nonumber \\
                                           & \phantom{\leq\;\;} 1 + n_0 + \mathcal{R}_{n_0}(n_1 - 2,n_2) + \mathcal{R}_{n_0}(n_1 - 1,n_2 - 1) \tag{evaluate $\mathcal{R}_{n_0}(n_1 - 1,n_2)$} \nonumber \\
                                           & \leq 1 + n_0 + f(n_0,n_1,n_2 - 1)\; + \nonumber \\
                                           & \phantom{\leq\;\;} 1 + n_0 + f(n_0,n_1-1,n_2 - 1) + \mathcal{R}_{n_0}(n_1 - 2,n_2) \tag{by the induction hypothesis} \nonumber \\
                                           & \leq \mathcal{R}_{n_0}(0,n_2) + \sum_{i = 0}^{n_1 - 1} \left( 1 + n_0 + f(n_0,n_1 - i,n_2 - 1) \right). \tag{by performing these steps repeatedly}
    \end{align}
    So overall, we obtain
    \begin{align} \mathcal{R}_{n_0}(n_1,n_2) &
              \leq \mathcal{R}_{n_0}(0,n_2) + \sum_{i = 0}^{n_1 - 1} \left( 1 + n_0 + f(n_0,n_1 - i,n_2 - 1) \right) \tag{by \eqref{sumClaim}} \\
                                         & = n_0 + \sum_{i = 0}^{n_1 - 1} \left( 1 + n_0 + f(n_0,n_1 - i,n_2 - 1) \right) \nonumber \\
                                         & \leq n_0 + n_1 \cdot (1 + n_0 + f(n_0,n_1,n_2 - 1)) \tag{as $f(n_0,n_1 - i,n_2 - 1) \leq f(n_0,n_1,n_2 - 1)$}\nonumber \\
                                         & = n_0 + n_1 \cdot (1 + 2\cdot n_0) + (n_2 - 1) \cdot (1 + 2\cdot n_0) \cdot n_1^{n_2}\nonumber \\
                                         & \leq f(n_0,n_1,n_2) \label{thm:localRRF:proofEq}.
    \end{align}
    Here, \eqref{thm:localRRF:proofEq} holds as $n_1 \cdot (1 + 2\cdot n_0) \leq (1 + 2\cdot n_0 ) \cdot n_1^{n_2}$ for $n_2> 0$.

  \end{myproof}
}

\begin{example}
  \label{exa:localRB}
  With the first \rrf{} of \Cref{exa:rrf}, \Cref{thm:localRRF} yields the local runtime bound $\locParameter{t_1}{\set{t_1}}(\location_1) = \approx{r_{\mathrm{d}}(\location_1)}
    = x$ (since $r_{\mathrm{f}}(\location_1) = 0$). With the second \rrf{} of \Cref{exa:rrf}, we obtain $\locParameter{t_5}{\set{t_4,t_5}}(f_1) = \approx{r_{\mathrm{f}}(f_1)} \; \cdot \; \approx{r_{\mathrm{tf}}(f_1)}^{\approx{r_{\mathrm{f}}(f_1)}} = a$, since $r_{\mathrm{d}}(f_1) = 0$, $r_{\mathrm{tf}}(f_1) = 1$, and $r_{\mathrm{f}}(f_1) = a$.
\end{example}

\subsection{Lifting Local Runtime Bounds to Global Runtime Bounds}
\label{sec:lifting}
To lift local to global runtime bounds, we consider those transitions and function calls which start an evaluation of the subprogram $\TSet'$.
Remember that $\fun(\LSet_{\TSet'})$ denotes the set of all function calls $\location(\, \wildcard \,)\in\FSet$ with $\location\in\LSet_{\TSet'}$, where $\LSet_{\TSet'}$ are the start locations of the transitions in the subprogram $\TSet'$.
\begin{definition}[Entry Transitions and Function Calls]
  Let $\emptyset \neq \TSet' \subseteq \TSet$ be a non-empty set of transitions.
  Then $\entryT_{\TSet'} = \{r\in\TSet\setminus\TSet'\mid \exists\, \location\in\LSet_{\TSet'}.\, r = (\,\wildcard,\wildcard,\wildcard,\location)\}$ is the set of \emph{direct entry transitions} and $\entryF_{\TSet'} = \{r \in\TSet\setminus\TSet'\mid \fun(r) \cap \fun(\LSet_{\TSet'}) \neq \emptyset \}$ is the set of \emph{entry (function) calling transitions} for $\TSet'$.
\end{definition}

\begin{example}
  For \Cref{fig:ITS} and the subprogram $\TSet' = \{ t_1 \}$ with the locations $\LSet_{\{t_1\}} = \{ \location_1 \}$, we have the entry transitions $\entryT_{\{t_1\}}=\{ t_0 \}$.
  Moreover, considering $\TSet' = \{t_4,t_5\}$ yields $\LSet_{\{t_4,t_5\}} = \{ f_1 \}$ and $\entryF_{\{t_4,t_5\}} = \{ t_1 \}$.
\end{example}

\Cref{thm:LiftingRB} allows us to lift arbitrary local runtime bounds of a subprogram $\TSet'$ (e.g., local runtime bounds by \rrf{}s) to global runtime bounds for the full program with all transitions $\TSet$.
To this end, for every entry transition $\pret\in\entryT_{\TSet'} \cup \entryF_{\TSet'}$, we consider $\glo(\pret)$ to over-approximate how often a local run of $\TSet'$ is started.
In contrast to our previous work \cite{giesl2022ImprovingAutomaticComplexity,lommen2022AutomaticComplexityAnalysis,lommen2024TargetingCompletenessComplexity}, we also have to consider \emph{entry calling transitions} $r\in\entryF_{\TSet'}$.
Furthermore, we have to consider the size of the program variables after entering the subprogram by $\pret$ or by a function call $\rho$ in $\pret$.
Hence, we replace every program variable $v\in\VSet$ by its size bound $\Size(\vartheta,v)$ for $\vartheta = \pret$ or $\vartheta = \rho$, respectively.
This is denoted by ``$[v/\Size(\vartheta,v) \mid v\in\VSet]$''.
In \Cref{sect:SB}, we show how to infer size bounds automatically.
In the following, let $\location_r\in\LSet$ denote the target location of a transition $r$ and let $\location_\rho\in\LSet$ denote the location of a function call $\rho$, i.e., if $\rho = \location(\recupdate)$ then $\location_\rho = \location$.
Here, it is important to restrict ourselves to subprograms $\TSet'$ without \emph{initial}
transitions $(\location_0,\wildcard,\wildcard,\wildcard\,)$ that start in the initial location $\location_0$.
Let $\TSet_0$ denote the set of all initial transitions.
The reason is that initial transitions do not have predecessor (entry) transitions.
In fact, initial transitions always have the global runtime bound $1$ by construction since according to our definition of \its{s} in \Cref{def:recITS}, the initial location $\location_0$ cannot be reached by any transition.
\begin{restatable}[Lifting Local Runtime Bounds]{theorem}{thmLiftingRB}
  \label{thm:LiftingRB}
  Let $\glo$ be a (global) runtime bound, $\Size$ be a size bound, $\emptyset \neq \TSet' \subseteq \TSet\setminus \TSet_0$, and $t \in \TSet'$.
  Moreover, let $\locParameter{t}{\TSet'}$ be a local runtime bound for the transition $t$ w.r.t.\ $\TSet'$.
  Then $\glopr$ is also a global runtime bound, where $\glopr(t') = \glo(t')$ for all $t'\neq t$ and
  \[
    \begin{array}{rllll}
      \glopr(t)\; =\hspace*{2.7cm} \sum_{\pret\,\in\,\entryT_{\TSet'}} \;\glo(\pret) & \cdot            &
      \locParameter{t}{\TSet'}(\location_\pret) \, [v/\Size(\pret,v)                 & \mid v\in\VSet]  & \vspace{0.15cm} \\
      +\; \sum_{\pret\,\in\,\entryF_{\TSet'}}\;\sum_{\rho\,\in\, \fun(\pret) \,\cap\,
      \fun(\LSet_{\TSet'})} \;\glo(\pret)                                            & \cdot            &
      \locParameter{t}{\TSet'}(\location_\rho) \, [v/\Size(\rho,v)                   & \mid v\in\VSet]. &
    \end{array}
  \]
\end{restatable}
\makeproof{thm:LiftingRB}{
  \thmLiftingRB*
  \begin{myproof}
    We show that for all $t\in\TSet$, all $\initial\in\Valuation$, and all trees $\tree$ with $\tree_{\initial} \step^* \tree$, we have
    \[
      \eval{\glopr(t)}{\abs{\initial}} \geq \abs{\tree}_{\set{t}}.
    \]
    The case $t'\neq t$ is trivial, since $\glopr(t') = \glo(t')$ and $\glo$ is a runtime bound.
    For $t$, we have to show that
    \begin{align*}
      \eval{\glopr(t)}{\abs{\initial}} {} = {} & \Biggl\llbracket\sum_{\pret\,\in\,\entryT_{\TSet'}} \glo(\pret) \cdot \locParameter{t}{\TSet'}(\location_\pret) \, [v/\Size(\pret,v) \mid v\in\VSet] \\
      + \sum_{\pret\,\in\,\entryF_{\TSet'}}    & \hspace{0.15cm}\sum_{\rho\,\in\, \fun(\pret) \,\cap\,\fun(\LSet_{\TSet'})} \glo(\pret) \cdot
      \locParameter{t}{\TSet'}(\location_\rho) \, [v/\Size(\rho,v) \mid v\in\VSet]\Biggr\rrbracket_{\abs{\initial}} \\
                                               & {} \hspace*{-1.5cm} \geq {} \abs{\tree}_{\set{t}}.
    \end{align*}
    Let $\tree_1, \ldots, \tree_m$ be the maximal subtrees of $\tree$ where all edges are labeled with transitions from $\TSet'$ or function calls from $\fun(\TSet')\cap\fun(\LSet_{\TSet'})$.
    So, the subtrees $\tree_1, \ldots, \tree_m$ are constructed by only using $\step_{\TSet'\cup\set{\varepsilon}}$-steps, i.e., if $\tree_i$'s root node is labeled with $(\widetilde{\location_{i}},\actstate_i)$, then we have $(\{(\widetilde{\location_{i}},\actstate_i)\},\emptyset) \step_{\TSet'\cup\set{\varepsilon}}^* \tree_i$.
    Let $\vartheta_i\in\TSet\cup\FSet$ be the transition or the function call in the label of the edge to $(\widetilde{\location_{i}},\actstate_i)$ in $\tree$, i.e., $\vartheta_i$ starts the evaluation of the subprogram $\TSet'$.
    Let $k_i$ be the number of edges labeled with $t$ in $\tree_i$ and let $\tree$ have $k$ edges labeled with $t$, i.e., $\abs{\tree_i}_{\set{t}} = k_i$ and $\abs{\tree}_{\set{t}} = k$.
    Then we have $\sum_{i = 1}^m k_i = k$.

    As $\Size$ is a size bound, we have $\eval{\Size(\vartheta_i, v)}{\abs{\initial}} \geq \abs{\actstate_i(v)}$ for all $v\in\VSet$.
    Hence, by the definition of local runtime bounds and as bounds are weakly monotonically increasing functions, we can conclude that
    \begin{equation}
      \label{gloThmHelp}
      \eval{\locParameter{t}{\TSet'}(\location_{\vartheta_i}) \left[v/\Size(\vartheta_i,v) \mid v \in
          \VSet \right]}{\abs{\initial}} \; \geq \; \eval{\locParameter{t}{\TSet'}(\location_{\vartheta_i})}{\abs{\actstate_i}}
      \; \geq \; k_i.
    \end{equation}

    Finally, we analyze how many maximal $\TSet'$-subtrees can be reached via some $\vartheta\in\TSet\cup\FSet$ in the full tree $\tree$.
    Every entry transition $\vartheta_i = \pret\in\entryT_{\TSet'}$ can occur at most $\eval{\glo(\pret)}{\abs{\initial}}$ times in the tree $\tree$, as $\glo$ is a global runtime bound.
    Similarly, every function call $\vartheta_i = \rho \in \fun(\pret)\cap\fun(\TSet')$ for an $\pret\in\entryF_{\TSet'}$ can occur at most $\eval{\glo(\pret)}{\abs{\initial}}$ times in the tree $\tree$.
    Thus, we have
    \begin{align*}
      \eval{\glopr(t)}{\abs{\initial}} {} = {} & \Biggl\llbracket\sum_{\pret\,\in\,\entryT_{\TSet'}} \glo(\pret) \cdot \locParameter{t}{\TSet'}(\location_\pret) \, [v/\Size(\pret,v) \mid v\in\VSet] \\
      + \sum_{\pret\,\in\,\entryF_{\TSet'}}    & \hspace{0.15cm}\sum_{\rho\,\in\, \fun(\pret) \,\cap\,\fun(\LSet_{\TSet'})} \glo(\pret) \cdot
      \locParameter{t}{\TSet'}(\location_\rho) \, [v/\Size(\rho,v) \mid v\in\VSet]\Biggr\rrbracket_{\abs{\initial}} \\
      {} \geq {}                               & \sum_{i=1}^m \;\eval{\locParameter{t}{\TSet'}(\location_{\vartheta_i})\left[v/\Size(\vartheta_i,v) \mid v \in \VSet \right]}{\abs{\initial}} \\
      {} \geq {}                               & \sum_{i=1}^m k_i \tag{by \eqref{gloThmHelp}} \\
      {} = {}                                  & k \\
      {} = {}                                  & \abs{\tree}_{\set{t}}.
    \end{align*}
  \end{myproof}
}

\begin{example}
  \label{Ex:Runtime Bounds}
  We now compute the remaining global runtime bounds for the \its{} of \Cref{fig:ITS}, see \Cref{exa:globalRB}.
  For the transitions $t_1$ and $t_5$, we had inferred the local runtime bounds $\locParameter{t_1}{\set{t_1}}(\location_1) = x$ and $\locParameter{t_5}{\set{t_4,t_5}}(f_1) = a$ in \Cref{exa:localRB}.
  \begin{itemize}
    \item Thus, we obtain $\glo(t_1) = \glo(t_0) \cdot \locParameter{t_1}{\set{t_1}}(\location_1)\,[x/\Size(t_0,x)] = x$ (with $\glo(t_0) = 1$ and $\Size(t_0,x) = x$) by \Cref{thm:LiftingRB}.
    \item Furthermore, we have $\glo(t_5) = \glo(t_1) \cdot \locParameter{t_5}{\set{t_4,t_5}}(f_1)\,[a/\Size(\rho_1,a)] = x^2$ (with $\glo(t_1) = x$ and $\Size(\rho_1,a) = x$) by \Cref{thm:LiftingRB}.
    \item Moreover, $\locParameter{t_4}{\set{t_4,t_5}}(f_1) = 1$ is a local runtime bound since $t_4$ can only be executed once in the subprogram $\set{t_4,t_5}$.
          Here, we get $\glo(t_4) = \glo(t_1) \cdot \locParameter{t_4}{\set{t_4,t_5}}(f_1) = x$.
    \item Finally, $\locParameter{t_3}{\set{t_3}}(\location_2) = \log_2(y) + 2$ is also a local runtime bound, which cannot be inferred by linear ranking functions but by our technique based on \emph{twn}-loops \cite{lommen2022AutomaticComplexityAnalysis,lommen2023TargetingCompletenessUsing,lommen2024TargetingCompletenessComplexity,frohn2020TerminationPolynomialLoops,hark2020PolynomialLoopsTermination}
          (see\report{ App.\ C}\paper{ \cite[App.\ C]{report}} for the detailed construction).
          Lifting this local bound by \Cref{thm:LiftingRB} yields the global bound $\glo(t_3) = \glo(t_2) \cdot \locParameter{t_3}{\set{t_3}}(\location_2)\,[y/\Size(t_2,y)] = \log_2(y + x\cdot x^x) + 2$ (with $\glo(t_2) = 1$ and $\Size(t_2,y) = y + x\cdot x^x$).
  \end{itemize}
  Thus, our modular approach allows us to consider individual subprograms separately, to use different techniques to compute their local bounds, and to combine these local bounds into a global bound afterwards.
\end{example}

\section{Modular Computation of Size Bounds}
\label{sect:SB}
We now introduce our modular approach to compute size bounds.
To this end, we extend the technique of \cite{brockschmidt2016AnalyzingRuntimeSize} to handle ITSs with function calls.
For every result variable $\rv{\vartheta,v}\in\RVSet =(\TSet\cup\FSet)\times\VSet$, we define a \emph{local size bound} $\SizeLoc(\vartheta,v)\in\NN[\VSet\cup\FSet]$.
So $\SizeLoc(\vartheta,v)$ is a polynomial over the program variables and function calls (which are treated like variables).
When instantiating every function call $\rho$ in $\SizeLoc(\vartheta,v)$ by the size of its result, then $\SizeLoc(\vartheta,v)$ must be a bound on the size of $v$ after a \emph{single evaluation} step with $\vartheta$ (where ``size'' again refers to the absolute value).
Thus, local size bounds are more restrictive than local runtime bounds, which consider arbitrary runs within the subprogram rather than being limited to a single evaluation step.
Here, the result of a function call $\rho$ can be obtained as soon as the evaluation of the call ends in a configuration $(\location_\rho, \valuation_\rho)$ with $\location_\rho \in \RetSet$.
Then the result of the call has the size $\abs{\valuation_\rho}(v_{\location_\rho})$.
\begin{definition}[Local Size Bound]
  $\SizeLoc:\RVSet\to\NN[\VSet\cup\FSet]$ is a \emph{local size bound} if for all $\rv{\vartheta,v}\in\RVSet$, all evaluations $(\location',\valuation') \to_\vartheta (\location,\valuation)$ with $\valuation \neq \bot$, and all evaluations $(\location',\valuation') \to_\rho \circ \to^* (\location_\rho,\valuation_\rho)$ starting with some $\rho\in\FSet$ such that $\location_\rho\in\RetSet$ and $\valuation_\rho \neq \bot$, we have $\abs{\valuation}(v) \leq \eval{\; \SizeLoc(\vartheta,v)\,[\rho \, / \, \abs{\valuation_\rho}(v_{\location_\rho}) \mid \rho\in\FSet] \;}{\abs{\valuation'}}$.
\end{definition}

For every result variable $\rv{t,v}\in\TSet\times\VSet$ with $t = (\,\wildcard, \wildcard, \update, \wildcard\,)$, in practice we essentially use $\SizeLoc(t,v) = \approx{\update(v)}$, e.g., $\SizeLoc(t_1,y) = y + \rho_1$ and $\SizeLoc(t_1,x) = \approx{x-1} = x+1$ for the program from \Cref{fig:ITS}.
However, due to the guard $x>0$ of the transition $t_1$, here we can obtain the more precise local size bound $\SizeLoc(t_1,x) = x$.
Similarly, we essentially use $\SizeLoc(\rho,v) = \approx{\recupdate(v)}$ for every result variable $\rv{\rho,v}\in\FSet\times\VSet$ with $\rho = \location(\recupdate)$.
So for \Cref{fig:ITS}, we would obtain $\SizeLoc(\rho_2,a) = \approx{a-1} = a + 1$.
However, due to the guard $a > 0$ of the transition $t_5$ whose update contains the function call $\rho_2$, we can again obtain the more precise bound $\SizeLoc(\rho_2, a) = a$.

Next we construct a \emph{result variable graph} (RVG) which represents the influence of result variables on each other.
Here, a node $\rv{t,v}$ or $\rv{\rho,v}$ represents the value of $v$ after evaluating the transition $t$ or applying the update $\recupdate$ for the function call $\rho = \location(\recupdate)$, respectively.
For any polynomial $p \in \NN[\VSet \cup \FSet]$, let $\actA(p) \subseteq \VSet \cup \FSet$ denote the set of \emph{active arguments} of the polynomial $p$, i.e., $x\in\actA(p)$ iff $x \in \VSet \cup \FSet$ is a program variable or a function call which occurs in $p$.
Furthermore, for $\vartheta \in \TSet \cup \FSet$ let $\pre(\vartheta)$ denote transitions or function calls that directly precede $\vartheta$, i.e., $\vartheta'\in\pre(\vartheta)$ iff there exists an evaluation tree which contains the path $\to_{\vartheta'}\circ\to_{\vartheta}$.
Then the RVG has all result variables $\RVSet$ as its nodes, and it has an RV-edge from $\rv{\vartheta',v'}$ to $\rv{\vartheta,v}$ whenever $\vartheta'\in\pre(\vartheta)$ and $v'\in\actA(\SizeLoc(\vartheta,v))$.
In practice, we use efficiently computable over-approximations for $\pre(\cdot)$.
\begin{definition}[Result Variable Graph (RV-Edges)]
  \label{def:RVG}
  An \emph{RVG} has the nodes $\RVSet$ and the RV-edges $\{(\rv{\vartheta',v'},\rv{\vartheta,v})\mid\vartheta'\!\in\!\pre(\vartheta), v'\!\in\!\actA(\SizeLoc(\vartheta,v))\}$.
\end{definition}

\begin{example}
  \Cref{fig:RVG} depicts a part of the RVG for the program of \Cref{fig:ITS}.
  \begin{figure}[t]
    \begin{center}
      \begin{tikzpicture}[->,>=stealth',shorten >=1pt,auto,state without output/.append style={rectangle,minimum size=0pt,draw=none}]
        \node[state] (q0) {$\rv{t_0,x}$};
        \node[state] (q3) [right of=q0, node distance=1.7cm] {$\rv{t_1,x}$};
        \node[state] (q6) [right of=q3, node distance=1.7cm] {$\rv{\rho_1,a}$};
        \node[state] (q6_h1) [below of=q6, node distance=1cm] {};
        \node[state] (q6_h2) [above of=q6, node distance=1cm] {};
        \node[state] (q7) [right of=q6_h1, node distance=1.7cm] {$\rv{t_4,a}$};
        \node[state] (q8) [right of=q6_h2, node distance=1.7cm] {$\rv{t_5,a}$};
        \node[state] (q1) [right of=q8, node distance=2.3cm] {$\rv{t_0,y}$};
        \node[state] (q2) [right of=q1, node distance=1.7cm] {$\rv{t_2,y}$};
        \node[state] (q5) [below of=q2, node distance=0.85cm] {$\rv{t_3,y}$};
        \node[state] (q4) [below of=q1, node distance=0.85cm] {$\rv{t_1,y}$};
        \node[state] (q11) [right of=q6, node distance=1.7cm] {$\rv{\rho_2,a}$};

        \draw (q0) edge node {} (q3);
        \draw (q3) edge node {} (q6);
        \draw (q0) edge [bend left = 40] node {} (q6);
        \draw (q6) edge node {} (q8);
        \draw (q6) edge node {} (q11);
        \draw (q11) edge node {} (q7);
        \draw (q11) edge node {} (q8);

        \draw (q3) edge [loop below] node {} (q3);
        \draw (q1) edge node {} (q4);
        \draw (q4) edge [loop below] node {} (q4);
        \draw (q8) edge [loop left,red,dashed] node {} (q8);
        \draw (q8) edge [red,dashed,bend left = 20] node {} (q4);
        \draw (q7) edge [red,dashed,bend right = 20] node [right,bend right = 70] {} (q4);
        \draw (q11) edge [loop right,looseness=4] node {} (q11);
        \draw (q7) edge [red,dashed, bend right = 80] node [right] {} (q8);

        \draw (q1) edge node {} (q2);
        \draw (q4) edge node {} (q2);
        \draw (q2) edge node {} (q5);
        \draw (q5) edge [loop below] node {} (q5);
      \end{tikzpicture}
    \end{center}
    \caption{Part of the RVG with RV-Edges and \textcolor{red}{$\RetSet$}-Edges for \Cref{fig:ITS}}
    \label{fig:RVG}
  \end{figure}
  In the figure, the black non-dashed edges are RV-edges.
  For instance, there is an edge $\rv{t_1,x}\to\rv{\rho_1,a}$ as the value of $x$ after transition $t_1$ influences the value of $a$ after the function call $\rho_1$.
  More precisely, $t_1 \in \pre(\rho_1) = \{t_0, t_1 \}$ and $x \in \actA(\SizeLoc(\rho_1,a)) = \actA(x) = \{ x \}$.
  Note that we do not have an edge from $\rv{\rho_1,a}$ to $\rv{t_4,a}$ as $\SizeLoc(t_4,a) = 1$.
  While the full RVG has additional non-trivial SCCs,\footnote{
    As usual, a \emph{strongly connected component} (SCC) is a maximal subgraph with a path from each node to every other node.
    An SCC is \emph{trivial} if it consists of a single node without an edge to itself.
  } we omitted them from \Cref{fig:RVG} as they have no impact on the runtime.
\end{example}
So far, the values of function calls have been disregarded.
We now address this issue by extending the RVG by a second set of edges.
To this end, we say that $\rv{t',v_{\location'}}$ is an \emph{$\RetSet$-predecessor}
of $(t,\rho)$ for a transition $t$ containing a function call $\rho$ if the transition $t'$ ends in a return location $\location'\in\RetSet$ which is reachable from a function call $\rho = \location(\,\wildcard\,) \in \fun(t)$, i.e., if there is an evaluation $\to_{\rho}
  (\location,\wildcard\,)\to\dots\to_{t'}
  (\location',\wildcard\,)$.
This means that after executing $t'$, the function call $\rho$ is ``finished'' and its result is obtained in the return variable $v_{\location'}$.
Thus, the $\RetSet$-predecessor $\rv{t',v_{\location'}}$ corresponds to the value of the function call $\rho$ which we did not handle in the RV-edges.
Let $\pre^{\RetSet}(t,\rho)$ denote the set of all $\RetSet$-predecessors of $(t,\rho)$.
Whenever $t = (\,\wildcard,\wildcard,\update,\wildcard\,)$ has the update $\update$ with a function call $\rho \in \fun(\update(v))$ and $\rv{t',v_{\location'}} \in \pre^{\RetSet}(t,\rho)$, then there is an $\Omega$-edge from $\rv{t',v_{\location'}}$ to $\rv{t,v}$ in the RVG.
As for $\pre(\cdot)$, we use efficiently computable over-approximations for $\pre^{\RetSet}(\cdot,\cdot)$.

\begin{definition}[Result Variable Graph ($\RetSet$-Edges)]
  In addition to \Cref{def:RVG}, an \emph{RVG} also has the $\RetSet$-edges $\{(\rv{t',v_{\location'}},\rv{t,v})\mid \rv{t',v_{\location'}}\in\pre^{\RetSet}(t,\rho), \; t = (\,\wildcard,\wildcard,\update,\wildcard\,), \; \rho\in\fun(\update(v)) \}$.
\end{definition}

\begin{example}
  Reconsider the RVG in \Cref{fig:RVG} which depicts a part of the RVG for the program of \Cref{fig:ITS}.
  The RVG has four $\RetSet$-edges which are red-dashed in \Cref{fig:RVG}.
  We have $\pre^{\RetSet}(t_1,\rho_1) = \{\rv{t_4, a}, \rv{t_5, a} \}$, since both $t_4$ and $t_5$ end in the return location $f_2\in\RetSet$ whose return variable is $v_{f_2} = a$.
  Moreover, $t_1$'s update of $y$ contains the function call $\rho_1= f_1(\recupdate_1)$, and there are evaluations $\to_{\rho_1} (f_1, \wildcard\,) \to_{t_4} (f_2, \wildcard\,)$ and $\to_{\rho_1} (f_1, \wildcard\,) \to_{t_5} (f_2, \wildcard\,)$.
  Thus, there are $\RetSet$-edges from both $\rv{t_4,a}$ and $\rv{t_5,a}$ to $\rv{t_1,y}$ in the RVG.
  So \Cref{fig:RVG} has the five non-trivial SCCs $\{ \rv{t_1, x} \}$, $\{ \rv{\rho_2, a} \}$, $\{\rv{t_1, y} \}$, $\{ \rv{t_3, y} \}$, and $\set{\rv{t_5,a}}$, where the latter forms a cycle with an $\RetSet$-edge.
\end{example}

We already developed powerful techniques to lift local to global size bounds for ITSs without function calls in \cite{brockschmidt2016AnalyzingRuntimeSize,lommen2023TargetingCompletenessUsing,lommen2024TargetingCompletenessComplexity}.
To extend these technique to ITSs \emph{with} function calls, we now introduce a corresponding approach to obtain global size bounds for different types of components of the RVG.
More precisely, we consider the SCCs of the RVG in topological order and lift the local size bounds for the result variables of each SCC to global size bounds.
First, we treat trivial SCCs of the RVG in \Cref{sect:SB_trivial}.
Here, local size bounds can be lifted by taking the global size bounds for its ``predecessors'' into account.
Afterwards, we handle non-trivial SCCs in \Cref{sect:SB_nontrivial}.
To lift local to global size bounds in this case, in addition to the global size bounds of the ``predecessors'', one also has to consider (global) runtime bounds, since non-trivial SCCs may be traversed repeatedly.

\subsection{Size Bounds for Trivial SCCs}
\label{sect:SB_trivial}

We start with computing size bounds for \emph{trivial} SCCs $\set{\rv{\vartheta,x}}$ in the RVG.
To this end, we present two techniques (in \Cref{lem:lemTrivialSB,lem:lemTrivialRecSB}).
\Cref{lem:lemTrivialSB} considers the case where $\vartheta\in\FSet$ or $\vartheta\in\TSet$ with an update $\update$ such that $\fun(\update(x)) = \emptyset$.
The case $\fun(\update(x)) \neq \emptyset$ is handled in \Cref{lem:lemTrivialRecSB}.
If $\vartheta$ is an initial transition, i.e., $\pre(\vartheta) = \emptyset$, then $\SizeLoc(\vartheta,x)$ is already a (global) size bound.
Otherwise, if $\pre(\vartheta) \neq \emptyset$, then \Cref{lem:lemTrivialSB} over-approximates the sizes of the variables in $\SizeLoc(\vartheta,x)$ by the size bounds corresponding to the preceding transitions.
To this end, every program variable $v$ in $\SizeLoc(\alpha)$ for $\alpha = \rv{\vartheta, x}$ is replaced by its size bound $\Size(\vartheta',v)$ for a predecessor $\vartheta'\in\pre(\vartheta)$ in the RVG.
This is again denoted by $\SizeLoc(\alpha) \, [v / \Size(\vartheta',v) \mid v\in\VSet]$.
Thus, as mentioned, the SCCs of the RVG should be handled in topological order such that finite global size bounds may already be available for the predecessors of $\vartheta$.
\begin{restatable}[Size Bounds for Trivial SCCs Without Function Calls]{theorem}{lemTrivialSB}
  Let $\Size$ be a size bound and $\set{\rv{\vartheta,x}}$ be a trivial SCC of the RVG such that $\vartheta\in\FSet$ or $\fun(\update(x)) = \emptyset$ for the update $\update$ of $\vartheta\in\TSet$.
  Then $\Size'$ is also a size bound where $\Size'(\alpha) = \Size(\alpha)$ for all $\alpha \neq\rv{\vartheta,x}$, and for $\alpha = \rv{\vartheta,x}$ we have \label{lem:lemTrivialSB}
  \[
    \Size'(\alpha) =
    \left\{
    \begin{array}{l@{\quad}l}
      \SizeLoc(\alpha),                                                                                          & \text{if $\pre(\vartheta) = \emptyset$} \\
      \max_{\vartheta'\,\in\,\pre(\vartheta)}\set{\SizeLoc(\alpha) \, [v / \Size(\vartheta',v) \mid v\in\VSet]}, & \text{otherwise.}                       \\
    \end{array}
    \right.
  \]
\end{restatable}
\makeproof{lem:lemTrivialSB}{
  \lemTrivialSB*
  \begin{myproof}
    Let $\set{\rv{\vartheta,x}}$ be a trivial SCC such that $\vartheta\in\FSet$ or $\fun(\update(x)) = \emptyset$ for the update $\update$ of $\vartheta\in\TSet$.
    Moreover, let $\initial \in \Sigma$ and $\tree_{\initial} \step^* \tree$ such that $\tree$ contains a path $(\location_0,\initial) \to \dots \to_\vartheta(\_,\valuation)$ with $\valuation\neq\bot$.
    We have to prove that
    \[\abs{\valuation}(x) \; \leq \;
      \eval{\Size'(\vartheta,x)}{\abs{\initial}}.
    \]

    We first consider the case $\pre(\vartheta) = \emptyset$ (i.e., $\vartheta$ is an initial transition from $\TSet_0$).
    Note that by our definition of \its{s}, $\location_0$ can neither be the target location of a transition nor evaluated after a function call.
    Thus, the path of the tree $\tree$ has the form $(\location_0,\initial)\to_\vartheta(\wildcard,\valuation)$.
    Hence, we have $\eval{\Size'(\vartheta,x)}{\abs{\initial}} = \eval{
        \SizeLoc(\vartheta,x)}{\abs{\initial}}
      \geq \abs{\valuation}(x)$.
    Note that w.l.o.g., we have $\SizeLoc(\vartheta,x)\in\ZZ[\VSet]$ by the requirement $\fun(\update(x)) = \emptyset$ for the update $\update$ of the transition $\vartheta$.

    Otherwise, if $\pre(\vartheta) \neq \emptyset$, then the path in $\tree$ has the form $(\location_0,\initial) \to \dots\to_{\tilde{\vartheta}}(\_,\tilde{\valuation})\to_\vartheta(\_,\valuation)$ for some $\tilde{\vartheta}\in\pre(\vartheta)$.
    By the definition of size bounds, we have $\eval{\Size(\tilde{\vartheta},v)}{\abs{\initial}} \geq \abs{\tilde{\valuation}}(v)$ for all $v\in\VSet$.
    Thus, we obtain
    \begin{align*}
      \eval{\Size'(\vartheta,x)}{\abs{\initial}} & =
      \eval{\max_{\vartheta'\,\in\,\pre(\vartheta)}\set{\SizeLoc(\alpha) \, [v / \Size(\vartheta',v) \mid v\in\VSet]}}{\abs{\initial}} \\
                                                 & \geq \eval{\SizeLoc(\alpha) \, [v / \Size(\tilde{\vartheta},v) \mid v\in\VSet]}{\abs{\initial}} \\
                                                 & \geq \eval{\SizeLoc(\alpha)}{\abs{\tilde{\valuation}}} \\
                                                 & \geq \abs{\valuation}(x)
    \end{align*}
  \end{myproof}
}

\noindent
Note that due to the requirement on $\rv{\vartheta,x}$ in \Cref{lem:lemTrivialSB}, w.l.o.g.\ $\SizeLoc(\vartheta,x)\in\NN[\VSet\cup\FSet]$ only contains program variables $\VSet$, but no variable from $\FSet$.

\begin{example}
  \label{Ex:SizeBounds}
  Reconsider \Cref{fig:ITS,fig:RVG}.
  We have $\Size(t_0,x) = \SizeLoc(t_0,x) = x$ by \Cref{lem:lemTrivialSB} as $\pre(t_0) = \emptyset$.
  Moreover, we obtain $\Size(t_4,a) = 1$ since $\SizeLoc(t_4,a) = 1$.
  In \Cref{exa:NonTrivialSBAdditive} we will show that $\Size(t_1,x) = x$ is a size bound.
  As $\SizeLoc(\rho_1,a) = x$ and $\pre(\rho_1) = \set{t_0,t_1}$, this implies
  \[
    \Size(\rho_1,a) = \max\{x[x / \Size(t_0,x)],x[x / \Size(t_1,x)]\} = x.
  \]
\end{example}

The following theorem handles trivial SCCs $\set{\alpha}$ for $\alpha = \rv{t,x}$ where $t$'s update for $x$ contains function calls $\rho$.
Hence, in contrast to \Cref{lem:lemTrivialSB}, we have to instantiate the function calls $\rho$ in $\SizeLoc(\alpha)$ by the size bounds for the transitions of $\pre^{\RetSet}(t,\rho)$, as they reach the corresponding return locations.
If some of these function calls $\rho$ do not reach a return location, then we can set the size bound for $\rv{t,x}$ to $0$, because then $\to_t$ only reaches configurations of the form $(\,\wildcard, \bot)$.
If $\pre(t) = \emptyset$ (i.e., $t$ does not have a RV-predecessor in the RVG) but every function call $\rho_1,\dots,\rho_n\in\fun(\update(x))$ has a predecessor (i.e., $\pre^{\RetSet}(t,\rho_i) \neq \emptyset$ for all $i$), then we replace every such function call $\rho_i$ by a size bound for the $\RetSet$-predecessor.
More precisely, we replace $\rho_i$ in $\SizeLoc(\alpha)$ by $\Size(\beta_i)$ for $\beta_i \in\pre^{\RetSet}(t,\rho_i)$.
In the following, this is denoted by $\SizeLoc(\alpha)\,[\rho_i / \Size(\beta_i) \mid i\in[n]]$ where $[n]$ is the set $\set{1,\dots,n}$.
Otherwise, if $t$ has RV-predecessors in the RVG, then we also have to instantiate the program variables by the size bounds corresponding to the preceding transitions, as in \Cref{lem:lemTrivialSB}.
Afterwards, the function calls $\rho_i$ are handled as in the previous case.
Note that the order of the substitutions is important here.
Replacing a function call $\rho_i$ with the size bound $\Size(\beta_i)$ \emph{before} substituting\linebreak
program variables by size bounds from the preceding transitions is unsound in general, as variables in $\Size(\beta_i)$ would then also be substituted incorrectly.
\begin{restatable}[Size Bounds for Trivial SCCs With Function Calls]{theorem}{lemTrivialRecSB}
  \label{lem:lemTrivialRecSB}
  Let $\Size$ be a size bound and $\set{\rv{t,x}}$ be a trivial SCC of the RVG such that $t\in\TSet$ and $\fun(\update(x)) \neq \emptyset$.
  Then $\Size'$ is also a size bound where $\Size'(\alpha) = \Size(\alpha)$ for all $\alpha\neq\rv{t,x}$, and for $\alpha = \rv{t,x}$ with $\fun(\update(x)) = \{\rho_1,\dots,\rho_n \}$, we have
  \[
    \Size'(\alpha) = \left\{
    \begin{array}{@{}l}
      0, \hspace*{\fill} \text{ if $\pre^{\RetSet}(t,\rho_i) = \emptyset$ for some $i\in[n]$\phantom{.}} \bigskip                         \\
      \max\limits_{\beta_i\,\in\,\pre^{\RetSet}(t,\rho_i)}\set{\SizeLoc(\alpha)\,[\rho_i / \Size(\beta_i) \mid i\in[n]]}, \vspace{-0.1cm} \\
      \hspace*{\fill} \text{if all $\pre^{\RetSet}(t,\rho_i) \neq \emptyset$ and $\pre(t) = \emptyset$\phantom{.}} \bigskip               \\
      \max\limits_{\substack{\beta_i\,\in\,\pre^{\RetSet}(t,\rho_i)                                                                       \\\vartheta'\,\in\,\pre(t)}}\set{\SizeLoc(\alpha)\, [v / \Size(\vartheta',v)\mid v \in \VSet] \, [\rho_i / \Size(\beta_i)\mid i \in [n]]}, \vspace{-0.45cm}\\
      \hspace*{\fill} \text{if all $\pre^{\RetSet}(t,\rho_i) \neq \emptyset$ and $\pre(t) \neq \emptyset$.}
    \end{array}
    \right.
  \]
\end{restatable}
\makeproof{lem:lemTrivialRecSB}{
  \lemTrivialRecSB*
  \begin{myproof}
    Let $\set{\rv{t,x}}$ be a trivial SCC of the RVG such that $t\in\TSet$ and $\fun(\update(x)) = \{\rho_1,\dots,\rho_n \} \neq \emptyset$ for the update $\update$ of $t$.
    Moreover, let $\initial \in \Sigma$ and $\tree_{\initial} \step^* \tree$ such that $\tree$ contains a path $(\location_0,\initial) \to \dots \to_\vartheta(\_,\valuation)$ with $\valuation\neq\bot$.
    We have to prove that
    \[\abs{\valuation}(x) \; \leq \;
      \eval{\Size'(\vartheta,x)}{\abs{\initial}}.
    \]
    If $\pre^{\RetSet}(t,\rho_i) = \emptyset$ for some $i\in[n]$, then there are only evaluations $(\location_0,\initial) \to \dots \to_\vartheta(\_,\valuation)$ where $\valuation = \bot$.
    Hence, let $\pre^{\RetSet}(t,\rho_i) \neq \emptyset$ for all $i\in[n]$.

    We first consider the case $\pre(\vartheta) = \emptyset$ (i.e., $\vartheta$ is an initial transition from $\TSet_0$).
    Again, by our definition of \its{s}, $\location_0$ can neither be the target location of a transition nor evaluated after a function call.
    Thus, the path of the tree $\tree$ has the form $(\location_0,\initial)\to_t(\wildcard,\valuation)$.
    At the same time, $\tree$ also contains paths $(\location_0,\initial)\to_{\rho_i} (\location_i,\wildcard) \to \dots \to_{t_i'} (\location_i', \valuation_i')$ for all $i \in [n]$ where $\location_i'\in\RetSet$ since $\valuation \neq \bot$.
    Hence, $\rv{t_i',v_{\location_i'}} \in \pre^{\RetSet}(t,\rho_i)$.
    By the definition of size bounds, we have $\eval{\Size(t_i',v_{\location_i'})}{\abs{\initial}} \geq \abs{\valuation_i'}(v_{\location_i'})$.
    Hence, we obtain
    \begin{align*}
      \eval{\Size'(t,x)}{\abs{\initial}} & \geq \eval{\SizeLoc(t,x)\, [\rho_i / \Size(t'_i,v_{\location_i'}) \mid i\in[n]]}{\abs{\initial}} \\
                                         & \geq \eval{\SizeLoc(t,x)\, [\rho_i/\abs{\valuation_i'}(v_{\location_i'})\mid i\in[n]]}{\abs{\initial}} \\
                                         & \geq \abs{\valuation}(x).
    \end{align*}

    Otherwise, if $\pre(t) \neq \emptyset$, then the path in $\tree$ has the form $(\location_0,\initial)\to \dots\to_{\tilde{\vartheta}}(\tilde{\location},\tilde{\valuation})\to_t(\_,\valuation)$ for some $\tilde{\vartheta}\in\pre(t)$.
    At the same time, $\tree$ also contains paths $(\location_0,\initial)\to \dots\to_{\tilde{\vartheta}}(\tilde{\location},\tilde{\valuation}) \to_{\rho_i} (\location_i,\wildcard) \to \dots \to_{t_i'} (\location_i', \valuation_i')$ for all $i \in [n]$ where $\location_i'\in\RetSet$ since $\valuation \neq \bot$.
    By the definition of size bounds, we have $\eval{\Size(\tilde{\vartheta},v)}{\abs{\initial}} \geq \abs{\tilde{\valuation}}(v)$ for all $v\in\VSet$ and $\eval{\Size(t_i',v_{\location_i'})}{\abs{\initial}} \geq \abs{\valuation_i'}(v_{\location_i'})$ for all $i \in [n]$.
    Thus, for $\alpha = \rv{t,x}$ we obtain
    \begin{align*}
      \eval{\Size'(\alpha)}{\abs{\initial}} & \geq \eval{\SizeLoc(\alpha) \, [v /
          \Size(\tilde{\vartheta},v) \mid v\in\VSet] \; [\rho_i / \Size(t'_i,v_{\location_i'}) \mid
      i\in[n]]}{\abs{\initial}} \\
                                            & = \eval{\SizeLoc(\alpha) \, [v /
          \Size(\tilde{\vartheta},v) \mid v\in\VSet] \;
      [\rho_i / \eval{\Size(t'_i,v_{\location_i'})}{\abs{\initial}} \mid
      i\in[n]]}{\abs{\initial}} \\
                                            & \geq \eval{\SizeLoc(\alpha) \,
        [\rho_i / \abs{\valuation_i'}(v_{\location_i'}) \mid
      i\in[n]]}{\abs{\tilde{\valuation}}} \\
                                            & \geq \abs{\valuation}(x).
    \end{align*}
  \end{myproof}
}

\begin{example}
  Consider a variant of \Cref{fig:ITS} where we replace the update $\update(y)$ of $t_1$ by $\rho_1 = f_1(\recupdate_1)$.
  Thus, the self-loop at $\rv{t_1,y}$ is removed from the RVG in \Cref{fig:RVG}.
  Then, we can apply \Cref{lem:lemTrivialRecSB} on the trivial SCC $\set{\rv{t_1,y}}$.
  Assume that we already computed $\Size(t_4,a) = 1$ and $\Size(t_5,a) = x^x$ (see \Cref{Ex:SizeBounds,exa:NonTrivialSB}).
  We have $\pre(t_1) = \{ t_0,t_1 \}$, but $\SizeLoc(t_1,y) = \rho_1$ does not contain variables from $\VSet$.
  Hence, we get $\Size(t_1,y) = \max\set{\rho_1[\rho_1 / \Size(t_4,a)], \rho_1[\rho_1 / \Size(t_5,a)]} = x^x$ as we have the $\RetSet$-predecessors $\rv{t_4,a}$ and $\rv{t_5,a}\in\pre^{\RetSet}(t_1,\rho_1)$.
\end{example}

\subsection{Size Bounds for Non-Trivial SCCs}
\label{sect:SB_nontrivial}

Finally, we introduce our approach to handle non-trivial SCCs.
Let $C \subseteq \RVSet$ be the nodes of such an SCC.
Our approach can only be applied to SCCs where for all $\alpha \in C$, there exist $e_\alpha\in\NN$ and $s_\alpha\in\NN[\VSet]$ such that
\begin{equation}
  \label{SLoc-Requirement}
  \textstyle
  \SizeLoc(\alpha) \leq s_\alpha \cdot (\; e_\alpha + \sum_{v\, \in \,
      \actA(\SizeLoc(\alpha))\setminus\actA(s_\alpha)} \; v \; )
\end{equation}
where ``$\leq$'' is interpreted pointwise (i.e., the inequation must hold for all instantiations of the variables by natural numbers).
Here, $s_\alpha$ captures the scaling behavior of $\SizeLoc(\alpha)$, e.g., it allows us to consider updates of the form $\update(x) = 2\cdot x$ or $\update(x) = a\cdot x$ for a variable $a \in \VSet$.
Note that in \cite{brockschmidt2016AnalyzingRuntimeSize}, only constant factors $s_\alpha$ were allowed.
Similarly, $e_\alpha$ captures the additive growth in updates like $\update(x) = 1 + x$.
For instance, we have $s_\alpha = a$, $e_\alpha = 0$, and $\actA(\SizeLoc(\alpha))\setminus\actA(s_\alpha) = \set{x}$ for the update $\update(x) = a\cdot x$.
The restriction \eqref{SLoc-Requirement}
is essential for our approach, since it allows us to apply an ``accumulation'' argument when lifting local to global size bounds.
Note that all linear updates satisfy \eqref{SLoc-Requirement}.
Thus, our approach is applicable to a wide range of programs in practice.
However, we cannot express all non-linear updates (e.g., the update $x^x$ cannot be handled).

We now also define $\pre$ and $\pre^\RetSet$ for result variables.
For $\alpha \in \RVSet$, $\pre(\alpha)$ ($\pre^\RetSet(\alpha)$) is the set of all result variables $\alpha'$ with an RV-edge ($\RetSet$-edge) from $\alpha'$ to\linebreak
$\alpha$ in the RVG.
Furthermore, for any result variable $\alpha$ in the SCC $C$, let $V_\alpha = \set{v\in\VSet\mid\exists\vartheta.\,\rv{\vartheta,v}\in\pre(\alpha)\cap C}$ consist of all variables $v$ with an RV-edge to $\alpha$\linebreak
in $C$, and similarly, $F_\alpha = \set{v\in\VSet\mid\exists t.\,\rv{t,v}\in\pre^\RetSet(\alpha)\cap C}$ are all variables with an $\RetSet$-edge to $\alpha$ in $C$.
Recall that $\actA(p)$ is the set of active arguments of the polynomial $p\in\NN[\VSet\cup\FSet]$.
Finally, for any $p \in \NN[\VSet \cup \FSet]$, let $\actV(p) = \actA(p) \cap \VSet$ be $p$'s \emph{active variables} and $\actF(p) = \actA(p) \cap \FSet$ be $p$'s \emph{active function calls}.

We first consider ``additive'' local size bounds only, and afterwards we generalize our method to handle ``multiplicative'' local size bounds as well.

\paragraph{Additive Local Size Bounds:}
We first consider local size bounds which are additive, i.e., where $s_\alpha = 1$ and $|V_\alpha| + |F_\alpha| \leq 1$.
Note that both of these requirements are necessary to prevent non-additive, exponential growth.

To consider the additive growth, we over-approximate the sizes of variables on incoming edges from outside the SCC $C$.
For example, consider the additive update $\update(x) = x + y$ of a simple loop where $y$ is not changed with a runtime bound $rb$.
In this example, the value of $y$ at the entry of the loop is repeatedly added to $x$\linebreak
in each iteration, i.e., we have $V_{\alpha} = \{ x \}$.
To capture such \emph{initial} values of an SCC of the RVG, we introduce the expressions $init_{\alpha}$ for RV-edges and $init_{\alpha}^\RetSet$ for $\RetSet$-edges which over-approximate the incoming values.
Let $init_{\alpha}(v) = \max\{\Size(\vartheta,v) \mid \exists \vartheta\in\TSet\cup\FSet.\, \rv{\vartheta,v}\in\pre(\alpha)\setminus C\}$ be a bound on the size of $v$ when entering $C$ via an RV-edge to $\alpha$.
Analogously, $init_{\alpha}^\RetSet(v) = \max\{\Size(t,v) \mid \exists t\in\TSet.\, \rv{t,v}\in\pre^\RetSet(\alpha)\setminus C\}$ is a bound on the size of $v$ when entering $C$ via an $\RetSet$-edge to $\alpha$.
To take the effect of function calls $\rho$ into account, we have to consider the return variables of the return locations reachable from $\rho$.
To ease the presentation, we assume that every function call $\rho$ always returns the same variable $v_\rho$, i.e., that $\rho$ cannot reach two return locations $\location_1, \location_2$ with $v_{\location_1} \neq v_{\location_2}$.\footnote{This could be generalized to function calls $\rho$ whose set of return variables $\retVar(\rho) = \{ v_{\location'}
    \mid \rho$ reaches $\location' \in \RetSet \}$ it not a singleton. Then in \eqref{addEquation}, instead of considering all $\rho \in \actF(\SizeLoc(\alpha))$ with $v_\rho \, \not\in \, F_\alpha$, one would have to consider all $v \in \retVar(\rho) \setminus F_\alpha$.}
The execution of $\alpha$'s transition or function call then means that the values of the variables in $V_\alpha$ or $F_\alpha$ can be increased by adding $init_{\alpha}(v)$ for all $v\in\actV(\SizeLoc(\alpha))\setminus V_\alpha$ (or, respectively, by adding $init_{\alpha}^\RetSet(v_\rho)$ for all $\rho\in\actF(\SizeLoc(\alpha))$ and $v_\rho\not\in F_\alpha$, where $v_\rho$ is the return variable associated to $\rho$) plus the constant $e_\alpha$.

Reconsider our example with the update $\update(x) = x + y$ where $V_\alpha = \{ x \}$.
After $n$ iterations of the loop, the value of $x$ is increased by $n \cdot init_\alpha(y)$.
Using the runtime bound $rb$ of this loop, our approach would over-approximate this increase by $rb \cdot init_\alpha(y)$.

In general, the increase of the variables in $V_\alpha$ or $F_\alpha$ by $init_{\alpha}(v)$ and $init_{\alpha}^\RetSet(v_\rho)$ is repeated $rb_\alpha$ times, where $rb_\alpha\! = \! \glo(t)$ if $\alpha\! =\! \rv{t,v}$ and $rb_\alpha = \sum_{t\in\trans(\rho)}\glo(t)$ if $\alpha = \rv{\rho,v}$, i.e., $rb_\alpha$ is a bound on how often $\alpha$'s transition or function call is evaluated during a program run.
Thus, the following expression over-approximates the additive size-change resulting from $\alpha$ (ignoring the growth resulting from $V_\alpha$ and $F_\alpha$ for now):
\begin{equation}
  \label{addEquation}
  \textstyle
  add(\alpha) = rb_\alpha\cdot (\; e_\alpha + \sum\limits_{\substack{v\, \in \, \actV(\SizeLoc(\alpha))\\ v\, \not\in \, V_\alpha}} init_{\alpha}(v) \hspace{0.3cm}
  +
  \sum\limits_{\substack{\rho\, \in \, \actF(\SizeLoc(\alpha))\\ v_\rho \, \not\in \, F_\alpha}} init_{\alpha}^\RetSet(v_\rho) \; )
\end{equation}

We now take the growth resulting from the variables in $V_\alpha$ or $F_\alpha$ into account.
Here, one has to consider the initial values of the variables in $V_\alpha$ and $F_\alpha$ before entering the SCC $C$.
This leads to
\[
  \textstyle
  add(\alpha) \; + \; \sum_{v\in V_{\alpha}} \; init_\alpha(v) \; + \; \sum_{v\in F_{\alpha}} \; init_{\alpha}^\RetSet(v).
\]
Since we only have to consider the initial values of the variables in $V_\alpha$ and $F_\alpha$, they are not multiplied with the runtime bound.
So in our example with the update $\update(x) = x + y$ and $V_{\alpha} = \{ x \}$, $init_\alpha(x)$ is not multiplied with the runtime bound $rb$ and we would obtain the size bound $init_\alpha(x) + rb \cdot init_\alpha(y)$.
The following \Cref{thm:thmNonTrivialSBAdditive} summarizes the previous observations and yields size bounds for additive SCCs.
To simplify the pre\-sentation, in contrast to \cite{brockschmidt2016AnalyzingRuntimeSize}, we do not consider transitions individually.
\begin{restatable}[Size Bounds for Non-Trivial Additive SCCs]{theorem}{thmNonTrivialSBAdditive}
  \label{thm:thmNonTrivialSBAdditive}
  Let $\Size$ be a size bound and $C$ be a non-trivial SCC of the RVG, where for all $\alpha \in C$, $\SizeLoc(\alpha)$ satisfies \eqref{SLoc-Requirement} for a suitable $e_\alpha$ and $s_\alpha = 1$, and moreover $|V_\alpha| + |F_\alpha| \leq 1$.
  Then $\Size'$ is also a size bound where $\Size'(\alpha) = \Size(\alpha)$ for all $\alpha \in \RVSet \setminus C$, and $\Size'(\alpha) = \Size'(C)$ for all $\alpha\in C$, where
  \[ \textstyle
    \Size'(C) = \sum_{\alpha\in C}
    ( add(\alpha) + \sum_{v\in V_{\alpha}} init_{\alpha}(v)
    + \sum_{v\in F_{\alpha}} init_{\alpha}^\RetSet(v))
  \]
\end{restatable}
\makeproof{thm:thmNonTrivialSBAdditive}{
  \thmNonTrivialSBAdditive*
  \begin{myproof}
    This claim can be proven analogously to \Cref{thm:thmNonTrivialSB} by setting $scale(\alpha) = 1$.
  \end{myproof}
}

\begin{example}
  \label{exa:NonTrivialSBAdditive}
  Reconsider \Cref{fig:ITS,fig:RVG}.
  We now infer size bounds for the non-trivial SCCs $\set{\rv{t_1, x}}$ and $\set{\rv{\rho_2,a}}$.
  \begin{itemize}
    \item For $\alpha =\rv{t_1, x}$ with $\SizeLoc(t_1,x) = x$, we have $s_\alpha = 1$, $e_\alpha = 0$, $V_\alpha = \set{x}$, $F_\alpha = \emptyset$, $\actV(\SizeLoc(t_1,x)) = \set{x}$, and $\actF(\SizeLoc(t_1,x)) = \emptyset$.
          This implies $add(\alpha) = 0$ and $init_\alpha(x) = \Size(t_0,x) = x$.
          Thus, we obtain the size bound $\Size(t_1, x) = x$.
    \item Similarly, for $\alpha = \rv{\rho_2,a}$ with $\SizeLoc(\rho_2,a) = a$, we obtain $s_\alpha = 1$, $e_\alpha = 0$, $V_\alpha = \set{a}$, $F_\alpha = \emptyset$, $\actV(\SizeLoc(\rho_2,a)) = \set{a}$, and $\actF(\SizeLoc(\rho_2,a)) = \emptyset$.
          Thus, we have $add(\alpha) = 0$ and $init_\alpha(a) = \Size(\rho_1, a) = x$ by \Cref{Ex:SizeBounds}.
          This yields the size bound $\Size(\rho_2,a) = x$.
  \end{itemize}
\end{example}
\paragraph{Multiplicative Local Size Bounds:}
Now, we consider local size bounds where $s_\alpha \neq 1$ or $|V_\alpha| + |F_\alpha| > 1$.
Again, let us introduce the essentially ideas for the size bound computations with a simple loop with the runtime bound $rb$.
In our example, we consider the update $\update(x) = \update(y) = x + y$ where $x,y\in V_\alpha$.
Then, both $x$ and $y$ grow with a factor of two.
A similar effect would be obtained for scaling factors $s_\alpha > 1$.
If $|V_\alpha| + |F_\alpha| > 1$, then each execution of $\alpha$'s transition or function call may multiply the value of a variable by $|V_\alpha| + |F_\alpha|$.
As for the additive growth, this multiplication must be performed $rb_\alpha$ times.
Thus, we obtain $(|V_\alpha| + |F_\alpha|)^{rb_\alpha}$ as the scaling factor caused by $V_\alpha$ and $F_\alpha$.
The scaling factor $s_\alpha$ can be handled similarly.
However, here we need to be careful as $s_\alpha$ might contain variables.
To this end, we over-approximate every variable in $s_\alpha$ by the size bound of the predecessors and ensure that the scaling factor is at least $1$.
This is captured by $scale(\alpha)$ with
\begin{equation}
  \label{eq:scale_alpha}
  \textstyle
  scale(\alpha) = (\, \max_{\rv{\vartheta,\,\wildcard}\,\in\,\pre(\alpha)}\set{1, \, s_\alpha \, [v / \Size(\vartheta,v)
        \mid v\in\VSet]}\cdot\left(\abs{V_\alpha} +
  \abs{F_\alpha}\right)\, )^{rb_\alpha}
\end{equation}
So for our example $\update(x) = \update(y) = x + y $, we infer the scaling factor $scale(\alpha) = 2^{rb}$.
Since $V_\alpha = \{x,y\}$, here we obtain the size bound $2^{rb}\cdot (init_\alpha(x) + init_\alpha(y))$ by multiplying the scaling factor with the initial values $init_\alpha(x) + init_\alpha(y)$.

The following theorem shows how to compute size bounds for non-trivial SCCs $C$ in general, by accumulating $scale(\alpha)$ and $add(\alpha)$ for all $\alpha\in C$.
To this end, we have to multiply $scale(\alpha)$ with the initial size bounds for $V_\alpha$ and $F_\alpha$ as in the example.
(The other initial size bounds are already covered in $add(\alpha)$.)
So, the scaling effect of several result variables in $C$ has to be multiplied whereas the additive size change for several result variables in $C$ has to be added.
\begin{restatable}[Size Bounds for Non-Trivial SCCs]{theorem}{thmNonTrivialSB}
  \label{thm:thmNonTrivialSB}
  Let $\Size$ be a size bound and $C$ be a non-trivial SCC of the RVG, where for all $\alpha \in C$, $\SizeLoc(\alpha)$ satisfies \eqref{SLoc-Requirement} for suitable $e_\alpha$ and $s_\alpha$.
  Then $\Size'$ is also a size bound where $\Size'(\alpha) = \Size(\alpha)$ for all $\alpha \in \RVSet \setminus C$, and $\Size'(\alpha) = \Size'(C)$ for all $\alpha\in C$, where
  \[ \textstyle
    \Size'(C) =
    \prod_{\alpha\in C} scale(\alpha) \; \cdot \;
    (\, \sum_{\alpha\in C}
    ( add(\alpha) + \sum_{v\in V_{\alpha}} init_{\alpha}(v)
    + \sum_{v\in F_{\alpha}} init_{\alpha}^\RetSet(v))\,)
  \]
\end{restatable}
\makeproof{thm:thmNonTrivialSB}{
  \thmNonTrivialSB*
  \begin{myproof}
    Let $C$ be a non-trivial SCC of the RVG, let $\initial \in \Sigma$, and $\tree_{\initial} \step^* \tree$ such that $\tree$ contains a path $(\location_0,\initial) \to \dots \to_\vartheta(\wildcard,\valuation)$ with $\valuation\neq\bot$.
    We have to prove that
    \[
      \abs{\valuation}(v) \leq \eval{\Size'(\vartheta,x)}{\abs{\initial}}.
    \]
    If $\rv{\vartheta,x}\not\in C$, then $\Size'(\vartheta,x) = \Size(\vartheta,x)$ is a size bound by definition.

    So let us consider $\alpha = \rv{\vartheta,x}\in C$.
    Note that $\vartheta$ cannot be an initial transition as there are no transitions or function calls leading back to the initial location $\location_0$ (i.e., then $C$ would not be a non-trivial SCC).
    Hence, there exists a predecessor\\[1pt] $\tilde{\vartheta}$ of $\vartheta$ in the path, i.e., the path has the form
    \[(\location_0,\initial)\to \dots \to_{\tilde{\vartheta}}(\tilde{\location},\tilde{\valuation})\to_{\vartheta}(\location,\valuation).\]

    Note that we must have $\abs{V_\alpha} + \abs{F_\alpha} > 0$ for all $\alpha \in C$ as $C$ is a non-trivial SCC of the RVG.
    Thus, as $\eval{\max_{\vartheta'\,\in\,\pre(\vartheta)}\set{1,\, s_\alpha \, [v / \Size(\vartheta',v) \mid v\in\VSet]}}{\abs{\initial}} \geq 1$ for all $\initial\in\Valuation$ and $\alpha \in C$, we also have
    \begin{equation}
      \label{eq:GTOne}
      \eval{scale(\alpha)}{\abs{\initial}} \geq 1 \text{ for all $\initial\in\Valuation$ and $\alpha \in C$.}
    \end{equation}
    We prove our claim by induction on the number $\abs{\tree}_{\mathcal{I}}$ where $\mathcal{I} = \set{\vartheta\mid\rv{\vartheta,v}\in C}$.
    \paragraph{Induction Base:}
    Here, we have $\abs{\tree}_{\mathcal{I}} = 1$ and thus $\rv{\tilde{\vartheta},v}\not\in C$ for all $v\in\VSet$.
    Note that we have
    \begin{align}
      \eval{init_\alpha(v)}{\abs{\initial}} & = \eval{\max\{\Size(\vartheta,v) \mid \exists \vartheta\in\TSet\cup\FSet.\, \rv{\vartheta,v}\in\pre(\alpha)\setminus C\}}{\abs{\initial}} \nonumber \\
                                            & \geq \eval{\Size(\tilde{\vartheta},v)}{\abs{\initial}} \nonumber \\
                                            & \geq \abs{\valuation(v)} \label{eq:SBProof1a}
    \end{align}
    for all $v\in\actV(\SizeLoc(\alpha))$ as $\rv{\tilde{\vartheta},v}\not\in C$.
    We extend $\fun(\cdot)$ to function calls by defining $\fun(\rho) = \emptyset$ for all $\rho \in \FSet$.
    Then for all function calls $\rho_i\in\fun(\vartheta)$, there is a path $(\location_0,\initial)\to \dots \to (\tilde{\location}, \tilde{\valuation}) \to_{\rho_i} \dots \to_{t_i} (\widetilde{\location_i},\widetilde{\valuation_i})$ in $\tree$ such that $\tilde{\location_i}\in\RetSet$.
    For all $\rho_i \in\actF(\SizeLoc(\alpha))$, we have
    \begin{align}
      \eval{init_\alpha^\RetSet(v_{\rho_i})}{\abs{\initial}} & = \eval{\max\{\Size(t,v_{\rho_i}) \mid
      \exists t\in\TSet.\, \rv{t,v_{\rho_i}}\in\pre^\RetSet(\alpha)\setminus C\}}{\abs{\initial}}
      \nonumber \\
                                                             & \geq
      \eval{\Size(t_i,v_{\rho_i})}{\abs{\initial}} \nonumber \\
                                                             & \geq
      \abs{\tilde{\valuation}_i(v_{\rho_i})} \label{eq:SBProof2a}
    \end{align}
    as $\rv{t_i,v_{\rho_i}}\not\in C$ since $|\tree|_{\mathcal I} = 1$.
    Thus, we have
    \begin{align*}
      \eval{\Size'(\vartheta,x)}{\abs{\initial}} & \geq \Biggl\llbracket scale(\alpha)\cdot
      \left(add(\alpha) + \sum_{v\in V_\alpha} init_\alpha(v) + \sum_{v \in F_\alpha} init_\alpha^\RetSet(v)\right)\Biggr\rrbracket_{\abs{\initial}} \tag{by \eqref{eq:GTOne}} \\
                                                 & \geq \eval{s_\alpha\,[v/\Size(\tilde{\vartheta},v)\mid v\in\VSet]}{\abs{\initial}}\cdot \left(e_\alpha + \sum_{\substack{v\,\in\,\actV(\SizeLoc(\alpha)) \\ v\,\not\in\, \actV(s_\alpha)}} \hspace{-0.8cm}\eval{init_\alpha(v)}{\abs{\initial}}\right. \\
                                                 & + \left.\sum_{\rho_i\,\in\,\actF(\SizeLoc(\alpha))}\hspace{-0.8cm}\eval{init_\alpha^\RetSet(v_{\rho_i})}{\abs{\initial}}\right) \\
                                                 & \geq
      \eval{s_\alpha}{\abs{\tilde{\valuation}}}\cdot \left(e_\alpha +
      \sum_{\substack{v\,\in\,\actV(\SizeLoc(\alpha)) \\ v\,\not\in\, \actV(s_\alpha)}} \hspace{-0.8cm}\abs{\tilde{\valuation}(v)} +
      \sum_{\rho_i \,\in\,\actF(\SizeLoc(\alpha))}\hspace{-0.8cm}\abs{\tilde{\valuation}_i(v_{\rho_i})}\right)
      \tag{by \eqref{eq:SBProof1a}, \eqref{eq:SBProof2a}, and as $\eval{\Size(\tilde{\vartheta},v)}{\abs{\initial}} \geq \abs{\tilde{\valuation}}(v)$ for all $v\in\VSet$}
      \\
                                                 & \geq \eval{\SizeLoc(\vartheta,x)\, [\rho_i/
            \abs{\tilde{\valuation}_i}(v_{\rho_i})\mid \rho_i\in\fun(\vartheta)]}{\abs{\tilde{\valuation}}}
      \tag{by \eqref{SLoc-Requirement}} \\
                                                 & \geq \abs{\valuation}(x).
    \end{align*}

    \paragraph{Induction Step:}
    We have $\abs{\tree}_{\set{t}} \leq \eval{\glo(t)}{\abs{\initial}}$ for all $t\in\TSet$ and $\abs{\tree}_{\set{\rho}} \leq \eval{\sum_{t\in\trans(\rho)}
        \glo(t)}{\abs{\initial}}$ for all $\rho\in\FSet$.
    Now for any $\overline{\alpha} = \rv{\overline{\vartheta},\wildcard}$, we define the following expressions:
    \[\begin{array}{rcl}
        add(\overline{\alpha},\tree)   & = & \abs{\tree}_{\set{\overline{\vartheta}}}\cdot\left(e_{\overline{\alpha}} +
        \sum\limits_{\substack{v\,\in\,\actV(\SizeLoc(\overline{\alpha}))                                               \\ v\,\not\in\, \actV(s_{\overline{\alpha}})\cup V_{\overline{\alpha}}}} \hspace{-0.3cm}init_{\overline{\alpha}}(v) +
        \sum\limits_{\substack{\rho\,\in\,\actF(\SizeLoc(\overline{\alpha}))
        \\ v_\rho\,\not\in\,
        F_{\overline{\alpha}}}}\hspace{-0.3cm} init_{\overline{\alpha}}^\RetSet(v_\rho)\right)                          \\[7ex]
        scale(\overline{\alpha},\tree) & = &
        (\max_{\rv{\vartheta',\,\wildcard}\,\in\,\pre(\overline{\alpha})}\set{1, \,
          s_{\overline{\alpha}}\,[v/\Size(\vartheta',v)\mid
              v\in\VSet]} \cdot (\abs{V_{\overline{\alpha}}} +
        \abs{F_{\overline{\alpha}}}))^{\abs{\tree}_{\set{\overline{\vartheta}}}}
      \end{array}\]
    So compared to $add(\alpha)$ and $scale(\alpha)$, we have replaced the over-approximation of the runtime bound $rb_\alpha$ by concrete values.
    Let $\tree'$ be the tree which results from $\tree$ by removing $(\location,\valuation)$.
    Then we define
    \begin{align*}
      \Psi = & \prod_{\overline{\alpha}\in C_\vartheta} scale(\overline{\alpha},\tree')\cdot\hspace{-0.3cm}\prod_{\overline{\alpha}\in C\setminus C_\vartheta} scale(\overline{\alpha},\tree)\cdot\left(\sum_{\overline{\alpha}\in C_\vartheta} add(\overline{\alpha},\tree') + \sum_{\overline{\alpha}\in C\setminus C_\vartheta} add(\overline{\alpha},\tree) + \right. \\
             & \left.\sum_{\overline{\alpha}\in C}\left(\sum_{v\in V_{\overline{\alpha}}} init_{\overline{\alpha}}(v) + \sum_{v\in F_{\overline{\alpha}}} init_{\overline{\alpha}}^\RetSet(v)\right)\right)
    \end{align*}
    where $C_\vartheta = \set{\rv{\vartheta,v}\in C\mid v\in\VSet}$, i.e., $C_\vartheta$ contains all
    result variables of $C$ that have $\alpha$'s transition or function call $\vartheta$.
    For all $v \in V_\alpha \cup F_\alpha$, we have $\eval{\Psi}{\abs{\initial}} \geq \abs{\tilde{\valuation}(v)}$ by the induction hypothesis.
    Furthermore, for $v\in\actV(\SizeLoc(\alpha))\setminus V_\alpha$, we have $\rv{\tilde{\vartheta},v}\not\in C$ and $\rv{\tilde{\vartheta},v} \in \pre(\alpha)$, and thus
    \begin{align}
      \eval{init_\alpha(v)}{\abs{\initial}} & = \eval{\max\{\Size(\vartheta,v) \mid \exists \vartheta\in\TSet\cup\FSet.\, \rv{\vartheta,v}\in\pre(\alpha)\setminus C\}}{\abs{\initial}} \nonumber \\
                                            & \geq \eval{\Size(\tilde{\vartheta},v)}{\abs{\initial}} \nonumber \\
                                            & \geq \abs{\tilde{\valuation}(v)}. \label{eq:SBProof1}
    \end{align}
    Again, there is a path $(\location_0, \initial) \to \cdots \to (\tilde{\location},
      \tilde{\valuation}) \to_{\rho_i} \dots \to_{t_i}
      (\tilde{\location_i},\tilde{\valuation_i})$ in $\tree$ for each function call
    $\rho_i\in\fun(\vartheta)$ such that $\tilde{\location_i}\in\RetSet$.
    Similarly, for all $v_{\rho_i}\in \VSet\setminus F_\alpha$ with
    $\rho_{i}\in\actF(\SizeLoc(\alpha))$, we have $\rv{t_i,v_{\rho_i}}\not\in
      C$ and $\rv{t_i,v_{\rho_i}} \in \pre^\RetSet(\alpha)$. Thus,
    \begin{align}
      \eval{init_\alpha^\RetSet(v_{\rho_i})}{\abs{\initial}} & = \eval{\max\{\Size(t,v_{\rho_i}) \mid \exists t\in\TSet.\, \rv{t,v_{\rho_i}}\in\pre^\RetSet(\alpha)\setminus C\}}{\abs{\initial}} \nonumber \\
                                                             & \geq \eval{\Size(t_i,v_{\rho_i})}{\abs{\initial}} \nonumber \\
                                                             & \geq
      \abs{\tilde{\valuation}_i(v_{\rho_i})}. \label{eq:SBProof2}
    \end{align}
    Finally, we define the following expression for all $\overline{\alpha} \in C$:
    \[
      \Phi_{\overline{\alpha}} = e_{\overline{\alpha}} +
      \sum\limits_{\substack{v\,\in\,\actV(\SizeLoc(\overline{\alpha})) \\ v\,\not\in\,
          \actV(s_{\overline{\alpha}})\cup
          V_{\overline{\alpha}}}}\hspace{-0.3cm}init_{\overline{\alpha}}(v) +
      \sum_{\substack{\rho\,\in\,\actF(\SizeLoc(\overline{\alpha}))\\ v_\rho\,\not\in
          \, F_{\overline{\alpha}}}}\hspace{-0.6cm}init_{\overline{\alpha}}^\RetSet(v_\rho).
    \]
    To simplify the presentation, for $\overline{\alpha}$, let
    \[\widehat{s}_{\overline{\alpha}} = \max_{\rv{\vartheta',\,\wildcard}\,\in\,\pre(\overline{\alpha})}\set{1, \,
        s_{\overline{\alpha}} \, [v / \Size(\vartheta',v) \mid v\in\VSet]}.\]
    In particular, the following holds:
    \begin{equation}
      \label{eq:SBProof3}
      \eval{\widehat{s}_\alpha}{\abs{\initial}} \geq \eval{s_\alpha \, [v / \Size(\tilde{\vartheta},v) \mid v\in\VSet]}{\abs{\initial}} \geq \eval{s_\alpha}{\abs{\tilde{\valuation}}}.
    \end{equation}
    Thus, we now have
    \begin{align*}
      \eval{\Size'(\vartheta,x)}{\abs{\initial}} & = \Biggl\llbracket\prod_{\overline{\alpha}\in
      C} scale(\overline{\alpha}) \cdot \left(\sum_{\overline{\alpha}\in C} \left( add(\overline{\alpha}) \right.\right. \\
                                                 & \hspace{2cm} + \left.\left.\underbrace{\sum_{v\in V_{\overline{\alpha}}} init_{\overline{\alpha}}(v) +
        \sum_{v\in F_{\overline{\alpha}}}
        init_{\overline{\alpha}}^\RetSet(v)}_{init_{\overline{\alpha}}}\right)\right)
      \Biggr\rrbracket_{\abs{\initial}} \\
                                                 & \geq \Biggl\llbracket\prod_{\overline{\alpha}\in C_{\tilde{\vartheta}}} \widehat{s}_{\overline{\alpha}}\cdot \left(\abs{V_\alpha} + \abs{F_\alpha}\right) \cdot \prod_{\overline{\alpha}\in C_{\tilde{\vartheta}}} scale(\overline{\alpha},\tree') \cdot \prod_{\overline{\alpha}\in C\setminus C_{\tilde{\vartheta}}} scale(\overline{\alpha},\tree) \\
                                                 & \qquad \cdot \left(\sum_{\overline{\alpha}\in C_{\tilde{\vartheta}}} add(\overline{\alpha},\tree') + \sum_{\overline{\alpha}\in C_{\tilde{\vartheta}}} \Phi_{\overline{\alpha}}
      + \sum_{\overline{\alpha}\in C\setminus C_{\tilde{\vartheta}}} add(\overline{\alpha},\tree) \right. \\
                                                 & \left. \hspace{1cm} + \sum_{\overline{\alpha}\in C} init_{\overline{\alpha}}\right)\Biggr\rrbracket_{\abs{\initial}} \tag{extract last evaluation step} \\
                                                 & = \Biggl\llbracket\prod_{\overline{\alpha}\in C_{\tilde{\vartheta}}} \widehat{s}_{\overline{\alpha}}\cdot \left(\abs{V_{\overline{\alpha}}} + \abs{F_{\overline{\alpha}}}\right) \cdot \Psi + \prod_{\overline{\alpha}\in C_{\tilde{\vartheta}}} \widehat{s}_{\overline{\alpha}}\cdot \left(\abs{V_{\overline{\alpha}}} + \abs{F_{\overline{\alpha}}}\right) \\
                                                 & \qquad \cdot \prod_{\overline{\alpha}\in C_{\tilde{\vartheta}}} scale(\overline{\alpha},\tree') \cdot \prod_{\overline{\alpha}\in C\setminus C_{\tilde{\vartheta}}} scale(\overline{\alpha},\tree) \cdot \left(\sum_{\overline{\alpha}\in C_{\tilde{\vartheta}}} \Phi_{\overline{\alpha}} \right)\Biggr\rrbracket_{\abs{\initial}} \tag{by definition of $\Psi$} \\
                                                 & \geq \Biggl\llbracket\prod_{\overline{\alpha}\in C_{\tilde{\vartheta}}} \widehat{s}_{\overline{\alpha}}\cdot \left(\abs{V_{\overline{\alpha}}} + \abs{F_{\overline{\alpha}}}\right) \cdot \left(\Psi + \sum_{\overline{\alpha}\in C_{\tilde{\vartheta}}} \Phi_{\overline{\alpha}}\right)\Biggr\rrbracket_{\abs{\initial}} \tag{by \eqref{eq:GTOne}} \\
                                                 & \geq \eval{\widehat{s}_\alpha}{\abs{\initial}}\cdot \left(\abs{V_\alpha} + \abs{F_\alpha}\right) \cdot \left(\eval{\Phi_\alpha}{\abs{\initial}} + \eval{\Psi}{\abs{\initial}}\right) \tag{by \eqref{eq:GTOne}} \\
                                                 & \geq \eval{s_\alpha}{\abs{\tilde{\valuation}}}\cdot \left(\abs{V_\alpha} + \abs{F_\alpha}\right) \cdot \left(\eval{\Phi_\alpha}{\abs{\initial}} + \eval{\Psi}{\abs{\initial}}\right) \tag{by \eqref{eq:SBProof3}} \\
                                                 & \geq \eval{s_\alpha}{\abs{\tilde{\valuation}}}\cdot \left(\eval{\Phi_\alpha}{\abs{\initial}} + \left(\abs{V_\alpha} + \abs{F_\alpha}\right) \cdot \eval{\Psi}{\abs{\initial}}\right) \tag{as $\abs{V_\alpha} + \abs{F_\alpha} \geq 1$} \\
                                                 & \geq
      \eval{s_\alpha}{\abs{\tilde{\valuation}}}\cdot
      \left(\eval{\Phi_\alpha}{\abs{\initial}} + \sum_{v\in V_\alpha}
      \eval{\Psi}{\abs{\initial}} +
      \sum_{\substack{\rho\,\in\,\actF(\SizeLoc(\alpha)) \\v_\rho\,\in \, F_\alpha}}\hspace{-0.8cm}\eval{\Psi}{\abs{\initial}}\right) \\
                                                 & \geq
      \eval{s_\alpha}{\abs{\tilde{\valuation}}}\cdot
      \left(\eval{\Phi_\alpha}{\abs{\initial}} + \sum_{v\in V_\alpha}
      \abs{\tilde{\valuation}(v)} + \sum_{\substack{\rho_i\,\in\,\actF(\SizeLoc(\alpha)) \\v_{\rho_i}\,\in \, F_\alpha}}\hspace{-.7cm}\abs{\tilde{\valuation}_i(v_{\rho_i})}\right) \tag{by the induction hypothesis}\\
                                                 & =
      \eval{s_\alpha}{\abs{\tilde{\valuation}}}\cdot \left(e_\alpha + \sum_{v\in V_\alpha}
      \abs{\tilde{\valuation}(v)} + \sum\limits_{\substack{v\,\in\,\actV(\SizeLoc(\alpha))
      \\ v\,\not\in\, \actV(s_{\alpha})\cup
          V_{\alpha}}}\hspace{-0.7cm} \eval{init_\alpha(v)}{\abs{\initial}}
      \right. \\
                                                 & \left. + \sum_{\substack{\rho_i\, \in\, \actF(\SizeLoc(\alpha)) \\v_{\rho_i}\, \in \, F_\alpha}}\hspace{-0.8cm}\abs{\tilde{\valuation}_i(v_{\rho_i}))} + \sum_{\substack{\rho_i\,\in\,\actF(\SizeLoc(\alpha))\\ v_{\rho_i}\,\not\in\, F_\alpha}}\hspace{-0.8cm}\eval{init_\alpha^\RetSet(v_{\rho_i})}{\abs{\initial}}\right) \tag{by definition of $\Phi_\alpha$}\\
                                                 & \geq \eval{s_\alpha}{\abs{\tilde{\valuation}}}\cdot \left(e_\alpha + \sum\limits_{\substack{v\,\in\,\actV(\SizeLoc(\alpha)) \\ v\,\not\in\, \actV(s_{\alpha})}}\hspace{-0.4cm} \abs{\tilde{\valuation}(v)} + \sum_{\rho_i \,\in\,\actF(\SizeLoc(\alpha))}\hspace{-0.8cm}\abs{\tilde{\valuation}_i(v_{\rho_i})}\right) \tag{by \eqref{eq:SBProof1} and \eqref{eq:SBProof2}} \\
                                                 & \geq \eval{\SizeLoc(\vartheta,x)\, [\rho_i/
            \abs{\tilde{\valuation}_i}(v_{\rho_i})\mid
            \rho_i\in\fun(\vartheta)]}{\abs{\tilde{\valuation}}}
      \tag{by \eqref{SLoc-Requirement}} \\
                                                 & \geq \abs{\valuation}(x).
    \end{align*}
  \end{myproof}
}

\begin{example}
  \label{exa:NonTrivialSB}
  We consider \Cref{fig:ITS,fig:RVG} again
  and infer size bounds for the missing SCCs $\set{\rv{t_5,a}}$ and $\set{\rv{t_1,y}}$.
  \begin{itemize}
    \item For $\alpha = \rv{t_5,a}$ with $\SizeLoc(t_5,a) = a\cdot\rho_2$, we have $s_\alpha = a$, $e_\alpha = 0$, $V_\alpha = \emptyset$, $F_\alpha = \set{a}$, $\actV(\SizeLoc(t_5,a)) = \set{a}$, and $\actF(\SizeLoc(t_5,a)) = \set{\rho_2}$.
          As $\pre(t_5) = \{ \rho_1, \rho_2 \}$, we have
          \[
            scale(\alpha) = (\max \{1, \, a[a/\Size(\rho_1,a)], \, a[a/\Size(\rho_2,a)] \})^{\glo(t_5)} = x^{x^2}
          \]
          (with $\glo(t_5) = x^2$ by \Cref{Ex:Runtime Bounds} and $\Size(\rho_1,a) = \Size(\rho_2,a) = x$ when using the invariant $x>0$ and thus, $\max\{1,x\} = x$), $add(\alpha) = x^2\cdot 0 = 0$, and $init_\alpha^\RetSet(a) = \Size(t_4,a) = 1$ by \Cref{Ex:SizeBounds}.
          Hence, we obtain the size bound $\Size(t_5,a) = x^{x^2}$.
          Note that \eqref{eq:scale_alpha} could be improved by considering a local runtime bound instead of the global runtime bound $rb_\alpha$.
          A similar improvement is used in \cite[Thm.\ 34]{lommen2023TargetingCompletenessUsing}.
          Thus, to obtain simpler bounds for readability, we use the size bound $\Size(t_5,a) = x^x$.
    \item Finally, for $\alpha = \rv{t_1,y}$ with $\SizeLoc(t_1,y) = y + \rho_1$, we have $s_\alpha = 1$, $e_\alpha = 0$, $V_\alpha = \set{y}$, $F_\alpha = \emptyset$, $\actV(\SizeLoc(t_1,y)) = \set{y}$, and $\actF(\SizeLoc(t_1,y)) = \set{\rho_1}$.
          Thus, we obtain $scale(\alpha) = 1$, $add(\alpha)= \glo(t_1) \cdot init^\RetSet_\alpha(a) = x \cdot \max \{ \Size(t_4,a),\Size(t_5,a) \} = x\cdot x^x$ (with $\glo(t_1) = x$ by \Cref{Ex:Runtime Bounds}, $\Size(t_4,a) = 1$, and $\Size(t_5,a) = x^x$), and $init_\alpha(y) = \Size(t_0,y) = y$.
          Hence, we get the size bound $\Size(t_1,y) = y + x\cdot x^x$.
          This also implies $\Size(t_2,y) = y + x\cdot x^x$.
  \end{itemize}
\end{example}

\section{Conclusion, Evaluation, and Related Work}
\label{sect:conclusion}

In this paper we presented a novel framework for complexity analysis of integer programs with (possibly non-tail recursive) function calls.
We introduced a\linebreak
unified approach that alternates between inferring runtime and size bounds.

The approach is modular, since it handles subprograms separately and allows us to use different techniques to generate bounds for the respective subprograms.
To infer runtime bounds, in addition to techniques based on multiphase-linear ranking functions \cite{giesl2022ImprovingAutomaticComplexity,ben-amram2017MultiphaseLinearRankingFunctions,ben-amram2019MultiphaseLinearRankingFunctions}
and ``complete'' techniques for \emph{twn}-loops \cite{lommen2022AutomaticComplexityAnalysis,lommen2023TargetingCompletenessUsing,lommen2024TargetingCompletenessComplexity,frohn2020TerminationPolynomialLoops,hark2020PolynomialLoopsTermination,brockschmidt2016AnalyzingRuntimeSize}, we introduced the new class of $\rho$-ranking functions to handle subprograms with function calls in \Cref{sect:RB}.
To infer size bounds, we showed in \Cref{sect:SB} how to generalize the respective technique from \cite{brockschmidt2016AnalyzingRuntimeSize} to ITSs with function calls.
So in particular, our novel approach can compute size bounds for the return values of function calls, i.e., the results of function calls are taken into account when analyzing complexity.
We implemented our new approach in the open-source tool \tool{KoAT} such that it can now also analyze ITSs with (possibly recursive) function calls.

Of course, our approach and its implementation have several limitations.
Currently, \tool{KoAT} can only successfully analyze programs where the arguments in recursive calls decrease w.r.t.\ a linear polynomial ranking function.
Moreover, our approach via \rrf{}s can only infer polynomial or exponential bounds.
So for example, we cannot obtain precise bounds for recursive algorithms with logarithmic runtime.
(Our implementation only computes logarithmic bounds for subprograms that are \emph{twn}-loops \cite{lommen2024TargetingCompletenessComplexity}.) Another restriction is that we can only lift local size bounds of the form \eqref{SLoc-Requirement} to global size bounds.
In addition, \tool{KoAT} may fail to infer finite bounds for several other reasons, e.g., some examples would need stronger invariants to make our approach succeed.
A detailed discussion and a corresponding list of examples to demonstrate the limitations of our approach and its implementation can be found on our webpage \cite{webpage}.

In the following, we conclude by discussing related work and by providing an experimental evaluation of our approach using the implementation in \tool{KoAT}.

\subsection{Related Work}

As mentioned in the introduction, there exist many approaches to analyze complexity of programs automatically, e.g., \cite{ben-amram2017MultiphaseLinearRankingFunctions,albert2019ResourceAnalysisDriven,sinn2017ComplexityResourceBound,brockschmidt2016AnalyzingRuntimeSize,Flores-MontoyaH14,giesl2022ImprovingAutomaticComplexity,hoffmann2017AutomaticResourceBound,lopez18IntervalBasedResource,carbonneaux2015CompositionalCertifiedResource,albert2012CostAnalysisObjectoriented,lommen2024ControlFlowRefinementProbabilistic,pham2024RobustResourceBounds,hoffmann2012MultivariateAmortizedResource,albert2013InferenceResourceUsage,handley2019LiquidateYourAssets}.
However, only few of them focus on programs with recursion or function calls.
While we already discussed an extension to recursive ITSs in \cite{brockschmidt2016AnalyzingRuntimeSize}, here the return values of function calls were ignored.

Techniques for complexity analysis of \emph{term rewrite systems} \cite{baader_nipkow_1998} can handle (possibly non-tail) recursion, but standard TRSs do not support built-in types like integers.
However in \cite{FroCoS17}, \emph{recursive natural transition systems} with potential non-tail recursion were introduced in order to study the connection between complexity analysis for TRSs and our approach from \cite{brockschmidt2016AnalyzingRuntimeSize} for complexity analysis of ITSs.
Here, the idea is to summarize (and subsequently eliminate) subprocedures by approximating their runtime and size.
Thus, this approach does not benefit from techniques such as our new class of ranking functions which allows us to handle subprograms with function calls directly.

Instead of representing integer programs as ITSs, there are also techniques based on \emph{cost equation systems} which can express non-tail recursive integer programs as well, e.g., \cite{Flores-MontoyaH14,flores-montoya2016UpperLowerAmortized}, implemented in the tool \tool{CoFloCo}.
This approach analyzes program parts independently and uses linear invariants to compose the results, i.e., it differs significantly from our approach which can also infer non-linear size bounds.
Similarly, in the tool \tool{PUBS} \cite{albert2008AutomaticInferenceUpper,albert2013InferenceResourceUsage}, \emph{cost relations}
are analyzed which are a system of recursive equations that capture the cost of the program.
There are also numerous approaches for automatic resource analysis of functional programs, often based on amortized analysis (see \cite{DBLP:journals/mscs/HoffmannJ22} for an overview).
For example, an approach for automatic complexity analysis of \tool{OCaml} programs is presented in \cite{hoffmann2012MultivariateAmortizedResource,hoffmann2017AutomaticResourceBound}, which however has limitations w.r.t.\ modularity, see \cite{FroCoS17}, and is restricted to polynomial bounds.
There are also several approaches based on types, e.g., the resource consumption of \tool{Liquid Haskell} programs is encoded in a type system in \cite{handley2019LiquidateYourAssets}, but here bounds are not inferred automatically.
Another line of work automatically infers bounds from recursive programs by generating and solving recurrence relations, e.g., \cite{Dynaplex2021,DBLP:journals/tplp/RustenholzKCL24}.
In future work, it might be interesting to integrate such techniques within our framework, as they can infer bounds that go beyond polynomials and exponentials.

There also exist tools which analyze the runtime complexity of \tool{C}-code, e.g.,
\tool{Loopus} \cite{sinn2017ComplexityResourceBound} or \tool{MaxCore} \cite{albert2019ResourceAnalysisDriven} with \tool{CoFloCo} or \tool{PUBS} in the backend.
For \tool{KoAT}, we used \tool{Clang} \cite{clang} and \tool{llvm2kittel}
\cite{falke2011TerminationAnalysisPrograms} to transform pointer-free \tool{C} programs into ITSs, and to handle more general \tool{C} programs, we developed the framework \tool{AProVE (KoAT + LoAT)} \cite{lommen2025AProVEKoATLoAT}, which also participates in the annual \emph{Software Verification Competition} (\emph{SV-COMP}) \cite{svcomp}.
\tool{AProVE} can also prove termination of recursive \tool{C} programs.
But when analyzing complexity, all these approaches are limited to \tool{C} programs without recursion.
In the future, it would be interesting to use our novel approach for \its{s} to extend complexity analysis to recursive \tool{C} programs, in particular in our framework \tool{AProVE (KoAT + LoAT)}.
\subsection{Evaluation and Implementation}
We implemented our novel results and integrated them into our open-source tool \KoAT{} which also features powerful techniques for subprograms without function calls \cite{brockschmidt2016AnalyzingRuntimeSize,lommen2022AutomaticComplexityAnalysis,lommen2023TargetingCompletenessUsing,lommen2024ControlFlowRefinementProbabilistic,lommen2024TargetingCompletenessComplexity,giesl2022ImprovingAutomaticComplexity}.
In the beginning, \KoAT{} preprocesses the program, e.g., by extending the guards of transitions with invariants inferred by \tool{Apron} \cite{jeannet2009ApronLibraryNumerical}.
For all SMT problems (including the generation of \rrf{s}), \KoAT{} calls \tool{Z3} \cite{moura2008Z3SMTSolver}.

For our evaluation, we use the set of 838 ITSs for complexity analysis of integer programs from the \emph{Termination Problems Data Base} (\emph{TPDB}) \cite{TPDB} which is used in the annual \emph{Termination and Complexity Competition} (\emph{TermComp}) \cite{giesl2019TerminationComplexityCompetition}.
While the \emph{TPDB} contains a large collection of ITSs without function calls, up to now there does not exist any such standard benchmark set for ITSs with function calls.
Thus, to obtain a representative collection of typical recursive programs, we extended our set by 45 new examples.
To this end, we transformed 20 recursive \tool{C} programs of the \emph{TPDB} and of the collection used at \emph{SV-COMP} into our novel formalism of \its{s}.
Moreover, we included our leading example (\Cref{fig:ITS}) along with several variants.
Our benchmarks also contain \its{s} with nested function calls and algorithms with multiple recursive function calls (e.g., the naive implementation of the Fibonacci numbers).
Moreover, our set of examples includes recursive versions of insertion sort and selection sort (where lists were abstracted to their lengths), for which we can infer quadratic runtime bounds.
In particular, there are also benchmarks that depend on non-linear size bounds, e.g., an algorithm which computes the product of two numbers recursively and then uses this result in a subsequent loop.
The average size of our benchmarks is 26 lines, whereas the largest one has 1615 lines.
In contrast, the average size of those benchmarks where \tool{KoAT} infers a finite bound is 17 lines and the largest of them has 246 lines.

Our benchmark collection, a detailed description of every new benchmark, and also the detailed results of our evaluation can be found on our webpage \cite{webpage}.

To distinguish the original \tool{KoAT} implementation of \cite{brockschmidt2016AnalyzingRuntimeSize} from our re-imple\-mentation, we refer to the tool of \cite{brockschmidt2016AnalyzingRuntimeSize} as \tool{KoAT1} in the following.
We evaluated our novel version of \tool{KoAT} with the approach of the current paper against the tools \tool{KoAT1} and \tool{CoFloCo}, which can also handle certain forms of recursion.
To this end, we transformed our novel benchmark \its{s} manually into their formalism.
We do not compare with \tool{PUBS}, because as stated in \cite{domenech2019ControlFlowRefinementPartial} by one of the authors of \tool{PUBS}, \tool{CoFloCo} is strictly stronger than \tool{PUBS}.
\begin{table}[t]
  \renewcommand{\arraystretch}{1.1}
  \setlength{\tabcolsep}{2.75pt}
  \begin{center}
    \begin{tabular}{lcccccccccccc}
      \toprule       & $\landau(1)$ & $\landau(\log(n))$ & \multicolumn{2}{c}{$\landau(n)$} & \multicolumn{2}{c}{$\landau(n^2)$} & \multicolumn{2}{c}{$\landau(n^{>2})$} & \multicolumn{2}{c}{$<\omega$} & $\mathrm{AVG^+(s)}$ & $\mathrm{AVG(s)}$                             \\
      \cmidrule(lr){2-2}
      \cmidrule(lr){3-3}
      \cmidrule(lr){4-5}
      \cmidrule(lr){6-7}
      \cmidrule(lr){8-9}
      \cmidrule(lr){10-11}
      \cmidrule(lr){12-12}
      \cmidrule(lr){13-13}
      \tool{KoAT}    & 128          & 11                 & 281                              & (18)                               & 121                                   & (8)                           & 28                  & (4)               & 583 & (34) & 3.45 & 21.78 \\
      \tool{KoAT1}   & 132          & 0                  & 231                              & (15)                               & 108                                   & (4)                           & 16                  & (2)               & 499 & (24) & 0.64 & 8.06  \\
      \tool{CoFloCo} & 125          & 0                  & 234                              & (1)                                & 95                                    &                               & 10                   &  (1)              & 464 & (2)  & 3.19 & 16.38 \\
      \bottomrule
    \end{tabular}
  \end{center}
  \caption{Evaluation on the Collection of ITSs and \its{s}}
  \label{fig:CINT}
\end{table}
\Cref{fig:CINT} shows the results of our evaluation, where as in \emph{TermComp}, we used a timeout of 5 minutes per example.
All tools were run inside an Ubuntu Docker container on a machine with an AMD Ryzen 7 3700X octa-core CPU and $8 \, \mathrm{GB}$ of RAM.
The first entry in every cell denotes the number of benchmarks for which the tool inferred the respective bound, where we consider both the ITSs from the \emph{TPDB} and our new \its{} benchmarks.
The number in brackets only considers the 45 new \its{} benchmarks.
The runtime bounds inferred by the tools are compared asymptotically as functions which depend on the largest initial absolute value $n$ of all program variables.
So for example, \tool{KoAT} proved an (at most) linear runtime bound\linebreak
for $128 + 11 + 281 = 420$ benchmarks, i.e., for these examples \tool{KoAT} can show that $\rc(\initial)\in\landau(n)$ for all initial states $\initial \in \Valuation$ where $|\initial(v)| \leq n$ for all $v \in \VSet$.
Overall, this configuration succeeds on $583$ examples, i.e., ``$< \omega$'' is the number of examples where a finite bound on the runtime complexity could be computed by the tool within the time limit.
Moreover, our tool \tool{KoAT} is able to prove termination for $626$ benchmarks.
So the termination proof succeeds for $43$ additional examples since one does not have to construct actual runtime bounds and does not have to consider size bounds.
``$\mathrm{AVG^+(s)}$'' denotes the average runtime of successful runs in seconds, whereas ``$\mathrm{AVG(s)}$'' is the average runtime of all runs.

In our experiments, \tool{KoAT} was the most powerful tool for runtime complexity analysis on both classical ITSs and -- due to the novel contributions of this paper -- also on ITSs with recursive function calls.
Note that \tool{KoAT1} heuristically applies loop-unrolling which might eliminate loops with constant runtime.
For this reason, \tool{KoAT1} infers constant runtime bounds for slightly more benchmarks than \tool{KoAT}.
For all other complexity classes in \Cref{fig:CINT}, \tool{KoAT} finds more examples for the respective class than the two other tools.
Moreover, \tool{KoAT} is the only of the three tools which can also infer logarithmic bounds (due to the integration of dedicated analysis techniques for subprograms that are \emph{twn}-loops).
Nevertheless, the three tools are ``orthogonal'', i.e., for each tool there are examples where the tool provides a finite bound and the other two tools fail.

Note that the tool \tool{LoAT} \cite{LoAT_CADE23,LoAT_IJCAR22} is able to prove absence of finite runtime bounds for $231$ of the $883$ benchmarks.
Thus, \tool{KoAT} is able to infer finite complexity bounds for $89\%$ of all $883 - 231 = 652$ benchmarks where this is potentially possible.
Our experiments demonstrate that handling return values directly yields significantly more precise bounds than prior approaches that simply ignore them.
In particular, to our knowledge \KoAT{} is currently the only tool which can infer a finite runtime bound for the recursive ITS from our leading example (\Cref{fig:ITS}).
So the new contributions of the paper are crucial in order to extend automated complexity analysis to the setting of recursive programs.
Moreover, verification frameworks for other programming languages can now be extended to analyze complexity of recursive programs by using \tool{KoAT} as a backend.

\KoAT's source code, a binary, a Docker image, and details on our new benchmarks and our evaluation are available at our webpage \cite{webpage}:
\[\mbox{\url{https://koat.verify.rwth-aachen.de/function-calls}}\]
This website also contains details on our input format for \its{s} and a \emph{web interface} to run different configurations of \KoAT{} directly online.
In addition, we also provide an artifact \cite{benchmarks} with \KoAT's binary and Docker images in order to reproduce our experiments.

\printbibliography
\pagebreak
\appendix
\section{Local Runtime Bounds for \rrf{s} with Multiple Function Calls}
\label{appendix:RRF}

In \Cref{thm:localRRF}, we showed how to obtain local runtime bounds from \rrf{s}
provided that the update polynomials for variables contain at most one function call.
Now, we present the more general version of \Cref{thm:localRRF} with arbitrary many function calls.
To this end, for any transition $t$ let $\nfc_{\TSet'}(t)$ denote the \underline{\textbf{n}}umber of \underline{\textbf{f}}unction \underline{\textbf{c}}alls $\location(\recupdate)\in\fun(t)$ with $\location\in\LSet_{\TSet'}$, and let $\nfc(\TSet') = \max \set{\nfc_{\TSet'}(t) \mid t \in \TSet'}$.
If every path has at most $n_0$ edges labeled with the transition $t$ and $t\not\in \TSet' \cap \trans(\fun(\LSet_{\TSet'}))$, $n_1$ edges labeled with the recursive transitions from $\TSet' \cap \trans(\fun(\LSet_{\TSet'}))$, and $n_2$ edges labeled with recursive function calls, then $\mathcal{R}_{n_0}(n_1,n_2)$ over-approximates the number of $t$-edges in any $\TSet'$-evaluation tree, where $\mathcal{R}_{n_0}(n_1,n_2)$ is defined via the following recurrence:

\vspace*{-.3cm}

\[\mbox{\small
    $\mathcal{R}_{n_0}(n_1,n_2) = \left\{
      \begin{array}{ll@{\hspace*{-2.6cm}}r}
         & n_0,                                                                                        & \text{if $n_1 = 0$, $n_2 = 0$, or $\nfc(\TSet') = 0$} \\
         & 1 + n_0 + \mathcal{R}_{n_0}(n_1 - 1,n_2) + \nfc(\TSet')\cdot\mathcal{R}_{n_0}(n_1,n_2 - 1), & \text{otherwise}                                      \\
      \end{array}
      \right.$}
\]

\vspace*{-.2cm}

\begin{figure}
  \center
  \begin{tikzpicture}[->,>=stealth',shorten >=1pt,auto]
    \node[state, minimum size=0.5cm, node distance=4.5cm] (qInit) {$N_0$};
    \node[state,draw=none] (h1) [right of=qInit, node distance=1.75cm] {$\cdots$};
    \node[state, label=above:{\footnotesize $\mathcal{R}_{n_0}(n_1,n_2)$}, minimum size=0.5cm,right of=h1, node distance=1.75cm] (q0) {$N_1$};
    \node[state, minimum size=0.5cm, node distance=4cm] (q2) [right of=q0] {$N_2$};
    \node[state,draw=none] (q1) [right of=q2, node distance=1.5cm] {$\cdots$};
    \node[state, label=above:{\footnotesize $\mathcal{R}_{n_0}(n_1 - 1,n_2)$}] (q2a) [right of=q1, minimum size=0.5cm, node distance=1.5cm] {$N_3$};

    \node[state,draw=none] (q3) [below of=q0,node distance=1cm] {$\cdots$};
    \node[state,draw=none] (q4) [below of=q0,node distance=1.5cm] {};

    \node[state,label=below:{\footnotesize $\mathcal{R}_{n_0}(n_1,n_2 - 1)$}, minimum size=0.5cm] (q5) [left of=q4,node distance=1.2cm,yshift=-.5cm] {};
    \node[state,label=below:{\footnotesize $\mathcal{R}_{n_0}(n_1,n_2 - 1)$}, minimum size=0.5cm] (q6) [right of=q4,node distance=1.2cm,yshift=-.5cm] {};

    \draw (qInit) edge [below] node {\scriptsize $t$} (h1);
    \draw (h1) edge [below] node {\scriptsize $t$} (q0);
    \draw (q2) edge [below] node {\scriptsize $t$} (q1);
    \draw (q1) edge [below] node {\scriptsize $t$} (q2a);
    \draw (q0) edge [below] node {\scriptsize $\TSet' \cap \trans(\fun(\LSet_{\TSet'}))$} (q2);

    \draw (q0) edge node [align=center,left,yshift=-0.3cm,xshift=-0.15cm] {\scriptsize $\rho_1$} (q5);
    \draw (q0) edge node [align=center,right,yshift=-0.3cm,xshift=0.15cm] {\scriptsize $\rho_{\nfc(\TSet')}$} (q6);

    \tikzset{
      position label/.style={
          below = 3pt, text height = 1.5ex, text depth = 1ex
        }, brace/.style={
          decoration={brace, mirror}, decorate
        }
    }

    \node[state,draw=none,node distance=0.25cm] (qh1) [below right of=q2] {};
    \node[state,draw=none,node distance=0.25cm] (qh2) [below left of=q2a] {};
    \node[state,draw=none,node distance=0.25cm] (qh3) [below right of=qInit] {};
    \node[state,draw=none,node distance=0.25cm] (qh4) [below left of=q0] {};

    \draw [brace,-,shorten >=0pt] (qh1.south east) -- (qh2.south west) node [position label, pos=0.5] {$d_2$ steps};
    \draw [brace,-,shorten >=0pt] (qh3.south east) -- (qh4.south west) node [position label, pos=0.5] {$d_1$ steps};
  \end{tikzpicture}
  \caption{Illustration of $\mathcal{R}_{n_0}(n_1,n_2)$}
  \label{fig:illustration_recurrence_extended}
\end{figure}
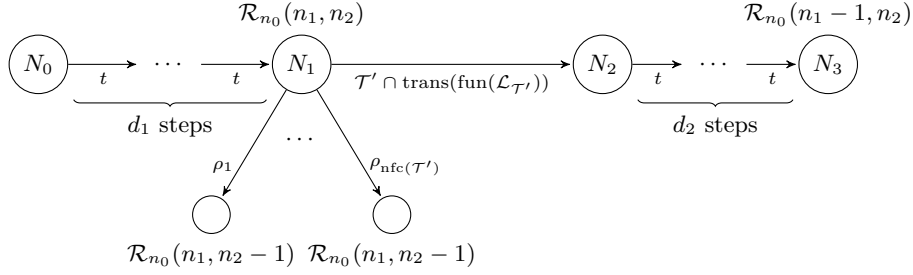

\noindent
This can be shown by induction on $n_1 + n_2$ (similar as in \Cref{sect:RB}):
If $n_1 = 0$, $n_2 = 0$, or $\nfc(\TSet') = 0$, then there is no recursive function call and thus, there can be at most $n_0$ edges labeled with the decreasing transition $t$.
The induction step is illustrated in \Cref{fig:illustration_recurrence_extended}.

Let us again first consider the case where $t$ is not a recursive transition.
Here, the path from the root node to the first node $N_1$ where a function is called recursively uses at most $d_1 \leq n_0$ edges labeled with $t$.
The node $N_1$ has at most $\nfc(\TSet')$ many outgoing edges labeled with recursive function calls and one outgoing edge to a node $N_2$ labeled with a recursive transition from $\TSet' \cap \trans(\fun(\LSet_{\TSet'}))$.
The function calls lead to at most $\nfc(\TSet')$ many subtrees where each contains at most $\mathcal{R}_{n_0}(n_1,n_2 - 1)$ many $t$-edges by the induction hypothesis.
The path from the node $N_2$ to the next node $N_3$ where a function might be called uses at most $d_2$ edges labeled with $t$, where we have $d_1 + d_2 \leq n_0$.
Finally, the subtree starting in node $N_3$ has at most $\mathcal{R}_{n_0}(n_1-1,n_2)$ many $t$-edges by the induction hypothesis.
Thus, the full tree has at most
\begin{align*}
  \abs{\tree}_{\{t\}} & \leq d_1 + \nfc(\TSet') \cdot \mathcal{R}_{n_0}(n_1,n_2 - 1) + d_2 + \mathcal{R}_{n_0}(n_1-1,n_2) \\
                      & \leq n_0 + \mathcal{R}_{n_0}(n_1 - 1,n_2) + \nfc(\TSet') \cdot \mathcal{R}_{n_0}(n_1,n_2 - 1)
\end{align*} many $t$-edges.

In the case where $t$ is recursive, as in \Cref{sect:RB} we obtain $\abs{\tree}_{\{t\}} \leq 1+ \mathcal{R}_{n_0}(n_1 - 1,n_2) + \nfc(\TSet') \cdot \mathcal{R}_{n_0}(n_1,n_2 - 1)$.
So in both cases, we have
\[ \abs{\tree}_{\{t\}} \leq 1 + n_0 + \mathcal{R}_{n_0}(n_1 - 1,n_2) + \nfc(\TSet')
  \cdot \mathcal{R}_{n_0}(n_1,n_2 - 1).\]

As in \Cref{app:proofs}, it can be shown that
\[n_0 + n_2 \cdot (1 + \left(1 + \nfc(\TSet')\right)\cdot n_0) \cdot (\nfc(\TSet')\cdot n_1)^{n_2}\] is an over-approximating closed form solution of $\mathcal{R}_{n_0}(n_1,n_2)$.
Hence, instantiating this closed form with the ranking functions yields the desired local runtime bound.
\begin{restatable}[Local Runtime Bounds by \rrf{}s]{theorem}{thmLocalRRF_extended}
  \label{thm:localRRF_extended}
  Let $\emptyset \neq \TSet'\subseteq\TSet$, let $t \in \TSet'$, and let $\langle r_{\mathrm{d}},r_{\mathrm{tf}},r_{\mathrm{f}} \rangle$ be a \rrf{} for $t$ w.r.t.\ $\TSet'$.
  Then $\loc:\LSet_{\TSet'}\to\BoundSet$ is local runtime bound for $t$ w.r.t.\ $\TSet'$, where for all $\location\in\LSet_{\TSet'}$, we define $\loc(\location)$ as:
  \[
    \approx{r_{\mathrm{d}}(\location)} \; + \; \approx{r_{\mathrm{f}}(\location)} \; \cdot \; \left(1 +
    \left(1 + \nfc(\TSet')\right)\cdot\approx{r_{\mathrm{d}}(\location)}\right) \; \cdot \; \left(\nfc(\TSet')\cdot
    \approx{r_{\mathrm{tf}}(\location)}\right)^{\approx{r_{\mathrm{f}}(\location)}}
  \]
\end{restatable}

\section{Local Runtime Bounds via \emph{twn}-Loops}
\label{appendix:RB}
In this appendix we briefly recapitulate how to infer runtime bounds for \emph{triangular weakly non-linear loops} (\emph{twn}-loops) based on our previous work \cite{lommen2022AutomaticComplexityAnalysis,lommen2023TargetingCompletenessUsing,lommen2024TargetingCompletenessComplexity,frohn2020TerminationPolynomialLoops,hark2020PolynomialLoopsTermination}.
This approach can be used to infer the local runtime bound $\locParameter{t_3}{\set{t_3}}(\location_2) = \log_2(y) + 2$ for transition $t_3$ in our leading example of \Cref{fig:ITS}.
This local runtime bound is needed in \Cref{Ex:Runtime Bounds} to compute $\glo(t_3)$.

An example for a terminating \emph{twn}-loop is:
\begin{equation}
  \label{WhileExample}
  \textbf{while } (x_2 - x_1 > 0 \, \wedge \, x_1 > 0) \textbf{ do }
  (x_1, x_2) \leftarrow (3\cdot x_1, \,	2\cdot x_2) \quad
\end{equation}
Note that this loop corresponds to transition $t_3$ ($x \cong x_1$ and $y \cong x_2$) with the additional invariant $x_1 > 0$.
In practice, \KoAT{} uses the tool \tool{Apron} \cite{jeannet2009ApronLibraryNumerical} to automatically infer such invariants.
Formally, a \emph{twn}-loop (over the variables $\vec{x} = (x_1,\ldots,x_\indv)$) is a tuple $(\guard,\update)$ with the guard $\guard$ and the update $\update: \VSet \rightarrow \ZZ[\VSet]$ for $\VSet = \{ x_1,\ldots,x_\indv \}$ such that for all $1\leq i \leq d$ we have $\update(x_i) = a_i\cdot x_i + p_i$ for some $a_i\in\ZZ$ and $p_i\in\ZZ[x_{1},\dots,x_{i-1}]$.
Thus, a \emph{twn}-update is triangular, i.e., the update of a variable does not depend on variables with higher indices.
Furthermore, the update is weakly non-linear, i.e., a variable does not occur non-linearly in its own update.

Our algorithm for the computation of runtime bounds for \emph{twn}-loops starts with computing closed forms for the loop update, which describe the values of the variables after $n$ iterations of the loop.
These closed forms can be represented as \emph{poly-exponential expressions}.
The set of all poly-exponential expressions is defined as $\PPEE = \{ \sum_{j=1}^k p_j \cdot n^{a_j} \cdot b_j^n \mid k, a_j\in \NN, \; p_j\in\QQ[\VSet], \; b_j\in\ZZ \}$.
\begin{example}
  \label{ex:closed form}
  The closed forms for the loop \eqref{WhileExample} are $\cl{x_1} = x_1 \cdot 3^n$ and $\cl{x_2} = x_2 \cdot 2^n$.
\end{example}

The following \Cref{lem:twn-poly} presents a construction based on closed forms which yields polynomial runtime bounds for terminating transitions $t = (\location,\guard,\update,\location)$ which correspond to \emph{twn}-loops.
We insert the closed forms of the update $\update$ into every atom $\alpha = p > 0$ of the guard $\guard$.
This results in a poly-exponential expression $pe_\alpha = \sum_{j=1}^{k_\alpha} p_{\alpha,j} \cdot n^{a_{\alpha,j}} \cdot b_{\alpha,j}^n\in\PPEE$ such that the summands are ordered w.r.t.\ the growth rate of $n^{a_{\alpha,j}} \cdot b_{\alpha,j}^n$.
Now, the polynomials $p_{\alpha,j}$ in $pe_\alpha$ determine the asymptotic complexity of the resulting local runtime bound.
\begin{theorem}[Polynomial Runtime Bounds for \emph{twn}-Loops]
  \label{lem:twn-poly}
  Let $t = (\location,\guard,\update,\location)$ be a terminating transition and for every atom $\alpha$ in $\guard$, let $pe_\alpha = \sum_{j=1}^{k_\alpha} p_{\alpha,j} \cdot n^{a_{\alpha,j}} \cdot b_{\alpha,j}^n\in\PPEE$ be a poly-exponential expression with $p_{\alpha,j} \neq 0$ for all $1 \leq j \leq k_\alpha$ and $(b_{\alpha,k_\alpha},a_{\alpha,k_\alpha}) \lex\ldots\lex (b_{\alpha,1},a_{\alpha,1})$ such that $pe_\alpha$ results from inserting the closed forms of $\update$ into $\alpha$.
  Then \[\locParameter{t}{\set{t}}(\location)
    = 2\cdot\max_{\alpha\text{ occurs in }\guard}\set{\approx{p_{\alpha,1}} + \dots +
      \approx{p_{\alpha,k_\alpha-1}}} + c\] is a local runtime bound where $c\in\NN$ is
  some computable constant.
\end{theorem}
\begin{example}
  \label{exa:poly}
  The loop \eqref{WhileExample} is terminating as the value of $x_1$ eventually outgrows the value of $x_2$.
  Inserting the closed forms of \Cref{ex:closed form} into the atoms yields $pe_{x_2 - x_1 > 0} = -x_1 \cdot 3^n + x_2 \cdot 2^n$ and $pe_{x_1 > 0} = x_1 \cdot 3^n$.
  So, we have $p_{x_2 - x_1 > 0, \; 1} = x_2$, $p_{x_2 - x_1 > 0, \;2}= -x_1$, and $p_{x_1> 0, \;1} = x_1$.
  Hence, for the transition $t$ and the location $\location$ corresponding to \eqref{WhileExample}, we obtain the polynomial local runtime bound $\locParameter{t}{\set{t}}(\location) = 2 \cdot \approx{p_{x_2 - x_1 > 0, \; 1}} + c = 2\cdot x_2 + c$ where $c = 1$ (see \cite{lommen2024TargetingCompletenessComplexity} for the detailed construction of $c$).
\end{example}

While \Cref{lem:twn-poly} always yields polynomial runtime bounds, we recently improved this to logarithmic runtime bounds if the exponential expressions are strictly decreasing, i.e., $b_{\alpha,k_\alpha} > \dots > b_{\alpha,1}$.
Intuitively, the reason is that then the summand $p_{\alpha,j} \cdot n^{a_{\alpha,j}} \cdot b_{\alpha,j}^n$ grows exponentially faster than all summands $p_{\alpha,i} \cdot n^{a_{\alpha,i}} \cdot b_{\alpha,i}^n$ for $i < j$.
\begin{theorem}[Logarithmic Runtime Bounds for \emph{twn}-Loops]
  \label{lem:twn-log}
  Let $t = (\location,\guard,\update,\location)$ be a terminating transition and for every atom $\alpha$ in $\guard$, let $pe_\alpha = \sum_{j=1}^{k_\alpha} p_{\alpha,j} \cdot n^{a_{\alpha,j}} \cdot b_{\alpha,j}^n\in\PPEE$ be a poly-exponential expression with $p_{\alpha,j} \neq 0$ for all $1 \leq j \leq k_\alpha$ and $b_{\alpha,k_\alpha} > \ldots > b_{\alpha,1}$ such that $pe_\alpha$ results from inserting the closed forms of $\update$ into $\alpha$.
  Then
  \[
    \locParameter{t}{\set{t}}(\location) = c'\cdot \log_2(\max_{\alpha\text{occurs in }\guard}\set{\approx{p_{\alpha,1}} + \dots + \approx{p_{\alpha,k_\alpha - 1}}}) + c
  \]
  is a local runtime bound where $c,c'\in\NN$ are some computable constants.
\end{theorem}
\begin{example}
  Reconsider \Cref{ex:closed form,exa:poly}:
  As the exponential terms in $pe_{x_2 - x_1 > 0} = -x_1 \cdot 3^n + x_2 \cdot 2^n$ and $pe_{x_1 > 0} = x_1 \cdot 3^n$ are strictly decreasing (i.e., $3 > 2$ for $pe_{x_2 - x_1 > 0}$), we can apply \Cref{lem:twn-log}.
  So for the transition $t$ and the location $\location$ corresponding to \eqref{WhileExample}, we obtain the \emph{logarithmic} local runtime bound $\locParameter{t}{\set{t}}(\location) = c'\cdot \log_2(\approx{p_{x_2 - x_1 > 0, \; 1}}) + c = c'\cdot \log_2(x_2) + c = \log_2(x_2) + 2$ where $c = 2$ and $c' = 1$ (see \cite{lommen2024TargetingCompletenessComplexity} for the detailed construction of $c$ and $c'$).
\end{example}

\end{document}